\documentclass{JHEP3}
\usepackage{amssymb}
\usepackage{amsfonts}
\usepackage{mathrsfs}

\newcommand{\e}{\begin{eqnarray}}
\newcommand{\ee}{\end{eqnarray}}

\newcommand{\CN}{{\cal N}}
\newcommand{\CL}{{\cal L}}

\def\a{\alpha}
\def\b{\beta}
\def\d{\delta}
\newcommand{\ep}{\epsilon}
\newcommand{\g}{\gamma}
\newcommand{\om}{\omega}
\newcommand{\p}{\psi}
\newcommand{\s}{\sigma}
\def\t{\tau}

\newcommand{\pa}{{a^{\prime}}}
\newcommand{\pb}{{b^{\prime}}}
\newcommand{\pc}{{c^{\prime}}}
\newcommand{\pd}{{d^{\prime}}}
\newcommand{\pe}{{e^{\prime}}}
\newcommand{\prf}{{f^{\prime}}}
\newcommand{\pg}{{g^{\prime}}}
\newcommand{\pri}{{i^{\prime}}}
\newcommand{\pj}{{j^{\prime}}}
\newcommand{\pk}{{k^{\prime}}}
\newcommand{\pl}{{l^{\prime}}}

\newcommand{\pu}{{u^{\prime}}}
\newcommand{\pv}{{v^{\prime}}}
\newcommand{\pw}{{w^{\prime}}}

\newcommand{\gap}{{\alpha^{\prime}}}
\newcommand{\gbp}{{\beta^{\prime}}}

\newcommand{\DA}{{\dot A}}
\newcommand{\DB}{{\dot B}}
\newcommand{\DC}{{\dot C}}
\newcommand{\DD}{{\dot D}}

\newcommand{\bi}{\bar{i}}
\newcommand{\bj}{\bar{j}}
\newcommand{\bk}{\bar{k}}
\newcommand{\bl}{\bar{l}}
\newcommand{\bm}{\bar{m}}
\newcommand{\bn}{\bar{n}}
\newcommand{\brp}{\bar{p}}
\newcommand{\bq}{\bar{q}}

\newcommand{\bt}{\bar{t}}
\newcommand{\bu}{\bar{u}}
\newcommand{\bv}{\bar{v}}
\newcommand{\bw}{\bar{w}}

\newcommand{\bp}{\bar{\psi}}

\newcommand{\hi}{\hat{i}}
\newcommand{\hj}{\hat{j}}
\newcommand{\hk}{\hat{k}}
\newcommand{\hl}{\hat{l}}

\newcommand{\tu}{\tilde{u}}
\newcommand{\tv}{\tilde{v}}

\title{Fusion of Superalgebras and $D=3, {\cal N}=4$
Quiver Gauge Theories}

\author{Fa-Min Chen$^{1}$, Yong-Shi Wu$^{2,3}$ \\
${}^1$ Fa-Min Chen \\
Department of Physics, Beijing Jiaotong Univeristy, Beijing 100044, China \\
${}^2$ Department of Physics and Center for Field Theory and Particle Physics,\\
Fudan University, Shanghai 200433, China\\
${}^2$ Department of Physics and Astronomy, University of Utah\\
Salt Lake City, UT 84112-0830, USA \\

E-mail: \email{fmchen@bjtu.edu.cn},
\email{wu@physics.utah.edu}}

\abstract{For further investigating the underlying structures of the $D=3, \CN=4$ Chern-Simons-matter (CSM) theories, we suggest a new concept and procedure
for  ``fusing" two superalgebras into a single new superalgebra.
The starting superalgebras may be those used in the previous
construction of the double-symplectic 3-algebras in the $\CN=4$ CSM theories: The bosonic
parts of these two superalgebras share at least one simple factor
or $U(1)$ factor.
We are able to provide two different methods to do the
``fusion". Several explicit examples are presented to demonstrate
the ``fusion" procedure.  We also generalize the ``fusion"
procedure so that more than two superaglebras can be fused into a
single one, provided some conditions are satisfied. It is shown that two or more $\CN=4$ theories with different gauge groups may be associated with the
same ``fused" superalgebra.}

\keywords{Fusion, Superalgebras, Symplectic 3-Algebras, Chern-Simons-matter
Theories}

\usepackage{bbm}
\usepackage{amsmath}
\usepackage{slashed}

\begin{document}

\section{Introduction} \label{Introduction}

In the recent years, the constructions of $D=3$, $\CN\geq 4$
superconformal Chern-Simons-matter (CSM) theories have attracted
lots of attention, because these theories are conjectured to be
the dual gauge theories of multiple M2-branes  \cite{Bagger}$-$\cite{Chen:pku1}. It has been demonstrated that general
Chern-Simons gauge theories with (or without) matter are
conformally invariant at the quantum level \cite{CSW1, CSW0, CSW2,
Piguet, Saemann1}. After incorporating the extended
supersymmetries into the CSM theories, these theories become
extended superconformal CSM theories, and we expect that they are
also conformally invariant at the quantum level.

The authors have been able to construct the $\CN=4$ quiver gauge
theory in terms of the double-symplectic 3-algebra \emph{or} the
$\CN=4$ three-algebra\footnote{The $\CN=4$ three-algebra is
obtained from the double-symplectic 3-algebra  by a
``contraction": the two structure constants $f_{a\pb c\pd}$ and
$f_{\pa b \pc d}$ are set to vanish, while the rest four structure
constants remain the same \cite{ChenWu3,ChenWu4}. However, in this
paper we focus on the double-symplectic 3-algebra
approach.}\cite{ChenWu3, ChenWu4}. (The double-symplectic
3-algebra is reviewed in Appendix \ref{secN4}.) The
double-symplectic 3-algebra consists of two sub symplectic
3-algebras. Denoting the generators of the two sub 3-algebras as
$T_a$ and $T_\pa$ ($a=1,\cdots,2R$ and $\pa=1,\cdots, 2S$),
respectively, then the generators of the double-symplectic
3-algebras are the disjoint union of the two sets of generators
$T_a$ and $T_\pa$. The untwisted multiplet $\Phi_A$ and the
twisted multiplet $\Phi^\prime_\DA$ of theory take values in these
two sub 3-algebras, respectively, i.e. $\Phi_A=\Phi^a_AT_a$ and
$\Phi^\prime_\DA=\Phi^\pa_{\DA} T_\pa$. Here $A=1,2$ and $\DA=1,2$
are fundamental indices of the $SU(2)\times SU(2)$ R-symmetry
group. The $\CN=4$ action can be built up by gauging \emph{part}
of the full symmetry generated by the double-symplectic 3-algebra.

Recently, using \emph{two} superalgebras $G$ and $G^\prime$ whose
bosonic parts share at least one simple factor or one $U(1)$
factor to construct the two sub symplectic 3-algebras, the authors
have been able to derive several classes $\CN=4$ theories with new gauge groups
and recover all known $\CN=4$ theories derived from ordinary Lie
(2-)algebra approach as well \cite{ChenWu4}. (The general forms of
$G$ and $G^\prime$ are presented in Appendix \ref{srt0}.)

In this paper we will propose the concept of ``fusing" two superalgebras $G$ and $G^\prime$ into a single closed
superalgebra, which is {\em not} a direct product of $G$ and
$G^\prime$, and will present two different methods to carry out
the fusion procedure. Of course, one of the motivations is that the
resulting superalgebra would be useful, or at least helpful, for
further investigating the underlying structures of the $\CN=4$ CSM theories, hopefully because of the
known close relationship between the superalgebras and the double-symplectic 3-algebras
in the $\CN=4$ CSM theories \cite{ChenWu4}. Also, the concept of fusing two superalgebras into a closed superalgebra and the problem of classifying these ``fused" superalgebras may be mathematically interesting.

More concretely, if we identify the generators of the two sub
3-algebras $T_a$ and $T_\pa$ with the fermionic generators of the
two superalgebras $Q_a$ and $Q_\pa$, respectively, \e T_a\doteq
Q_a\quad T_\pa\doteq Q_\pa, \ee then one may construct the
3-brackets in terms of double graded commutators on the
superalgebras \cite{ChenWu4}, for instances, \e\label{mxgnrts}
\quad [T_a, T_b; T_\pc]\doteq[\{Q_a, Q_b\}, Q_\pc],\quad [T_\pa,
T_\pb; T_c]\doteq[\{Q_\pa, Q_\pb\}, Q_c]. \ee Here $Q_a$ and
$Q_\pa$ are the fermionic generators of the two superalgebras $G$
and $G^\prime$, respectively; the double graded commutator
$[\{Q_a, Q_b\}, Q_\pc]$ is defined by an anticommutator and a
commutator. In order that the theory is physically interesting, we
must require that there are nontrivial interactions between the
twisted and untwisted multiplets; mathematically, we must require
that \e \quad [T_a, T_b; T_\pc]\neq0,\quad [T_\pa, T_\pb;
T_c]\neq0. \ee Taking account of (\ref{mxgnrts}), one is led to
\e\label{mxdgcm0} [\{Q_a, Q_b\}, Q_\pc]\neq0,\quad [\{Q_\pa,
Q_\pb\}, Q_c]\neq0. \ee
It has been proved that Eqs (\ref{mxdgcm0})
can be satisfied if the bosonic parts of $G$ and $G^\prime$ share
at least one simple factor or one $U(1)$ factor, provided that the
common part of bosonic parts of $G$ and $G^\prime$ is not a center
of $G$ and $G^\prime$ \cite{HosomichiJD,ChenWu4}. The first equation of
(\ref{mxdgcm0}) implies that there must be a nontrivial
anticommutator between $Q_a$ and $Q_\pc$ in the sense that
\e\label{nmxanti} \{Q_a, Q_\pc\}\neq 0, \ee \emph{provided that
the $Q_aQ_bQ_\pc$ Jacobi identity is obeyed}: \e\label{jacobi}
[\{Q_a, Q_b\}, Q_\pc]+[\{Q_a, Q_\pc\}, Q_b]+[\{Q_\pc, Q_b\},
Q_a]=0. \ee Actually, if $\{Q_a, Q_\pc\}=0$, the last two terms of
the RHS of (\ref{jacobi}) vanish. As a result, one must have
$[\{Q_a, Q_b\}, Q_\pc]=0$, which contradicts the first equation of
Eq. (\ref{mxdgcm0}).  We are led to Eq. (\ref{nmxanti}) again if
we combine the second equation of (\ref{mxdgcm0}) and the $Q_\pa
Q_\pb Q_c$ Jacobi identity. For some special cases, one can show
that the anticommutator (\ref{nmxanti}) does not vanish by
calculating it directly (Eqs. (\ref{mixcmmt}) and (\ref{qaqpb3})
are two examples).

On the other hand, according to the basic idea of supersymmetry,
an anticommutator of two fermionic generators gives a linear
combination of bosonic generators. It is therefore natural to
introduce a set of bosonic generators $M_{\tilde{u}}$ defined by
\e\label{ntrvanticm0}
\{Q_a,Q_\pc\}=t^{\tilde{u}}_{a\pc}M_{\tilde{u}}, \ee where
$t^{\tilde{u}}_{a\pc}$ are structure constants. Having defined
(\ref{ntrvanticm0}), if we also define $[M_{\tilde{u}}, M_{\tv}]$
and every commutator of $M_{\tilde{u}}$ and any generator of $G$
and $G^\prime$ properly, so that \emph{every Jacobi identity of
the ``total" superalgebra consisting of $M_{\tilde{u}}$ and all
generators of $G$ and $G^\prime$ is obeyed, we say that $G$ and
$G^\prime$ have been ``fused" into a single superalgebra which is
closed.} Note that one needs only to introduce the set of new
bosonic generators $M_{\tilde{u}}$ into the system for the purpose
of fusion; in particular, one does not have to introduce any new
fermionic generator into the system.

We demonstrate that if both $G$ and $G^\prime$ are orthosymplectic
or unitary superalgebras, and their bosonic parts of $G$ and
$G^\prime$ share at least one simple factor or $U(1)$ factor, we
can fuse them into a single superalgebra by using two distinct
methods. Some explicit examples are presented to demonstrate how
to construct this class of superalgebras by fusing two
superalgebras. For example, we can fuse $U(N_1|N_2)$ and
$U(N_2|N_3)$ into a single superalgbra $U(N_2|N_1+N_3)$. Here the
common bosonic part of $U(N_1|N_2)$ and $U(N_2|N_3)$ is $U(N_2)$.
Conversely, we can use the sub-superalgebras $U(N_1|N_2)$ and
$U(N_2|N_3)$ of $U(N_2|N_1+N_3)$ to construct the 3-algebra in the
$\CN=4$ theory, providing the bosonic parts of the two
sub-superalgebras share one common factor $U(N_2)$.



We are able to work out the general structure of the ``fused"
superalgebras by adding several new graded commutators into the
two superalgebras $G$ and $G^\prime$. We also generalize our
fusion procedure by showing that three or more superalgebras can
be fused into a single superalgebra by introducing more bosonic
generators (analogy to $M_{\tu}$ in (\ref{ntrvanticm0})).

The fusion procedure can be also generalized in another direction:
by adding some \emph{fermionic} generators into the system. We
will call this procedure a \emph{fermionic fusion} if one needs to
introduce at least a set of new fermionic generators (except for
new bosonic generators) for fusing two or more superalgebras into
a single one (see Sec. \ref{secmore2}).

We demonstrate that two or more theories with different gauge groups can be associated with the
same ``fused" superalgebra.


This paper is organized as follows. In section \ref{secfosps} and
section \ref{secfumun},  we present some explicit examples of the
new superalgebra fused by two superalgebras. The general structure
of the new superalgebra is worked out in section \ref{secGene}, and
some examples for fusing three or more superalgebras into a single
one are given as well.  We end section \ref{conclusions} with
conclusions and discussions. Several appendices are attached to
make the paper more self-contained. In Appendixes \ref{secN4} and
\ref{srt0}, we review the $\CN=4$ theories based on the 3-algebras
and the superalgebra realization of 3-brackets and fundamental identities (FIs),
respectively. We summarize our conventions in Appendix
\ref{Identities}. The commutation relations of some superalgebras
used to construct symplectic 3-algebras are given in Appendix
\ref{superalgebras}.


\section{Fusing Two Orthosympelctic Superalgebras }\label{secfosps}
As we explained in Section \ref{Introduction}, it is possible to
fuse two superalgebras whose bosonic parts share at least one
simple factor or $U(1)$ factor. In this section, we demonstrate
how to ``fuse" a pair of orthosympelctic superalgebras into a
single superalgebra by presenting two explicit examples.

\subsection{$Sp(2N_1)\times SO(N_2)\times Sp(2N_3)$
Gauge Group}\label{secexample} Here we choose the two
superalgebras as $G=OSp(N_2|2N_1)$ and $G^\prime=OSp(N_2|2N_3)$.
(The commutation relations of $OSp(M|2N)$ are given by Appendix
\ref{cmrsp22n}.) Namely the bosonic parts of the two superalgebras
share one simple factor $SO(N_2)$. We denote the fermionic
generators and the antisymmetric tensor of $OSp(N_2|2N_1)$ as \e
Q_a=Q_{\bi\hi}\quad {\rm and} \quad
\omega_{ab}=\omega_{\bi\hi,\bj\hj}=\d_{\bi\bj}\omega_{\hi\hj}, \ee
where $\bi=1,\cdots,N_2$ is an $SO(N_2)$ fundamental index, and
$\hi=1,\cdots, 2N_1$ an $Sp(2N_1)$ fundamental index. For
convenience, we cite the commutation relations of $OSp(N_2|2N_1)$
here
\begin{eqnarray}\label{OSpfus}
&&[M_{\bar i\bar j},M_{\bar k\bar
l}]=\d_{\bj\bk}M_{\bi\bl}-\d_{\bi\bk}M_{\bj\bl}+\d_{\bi\bl}M_{\bj\bk}
-\d_{\bj\bl}M_{\bi\bk},\nonumber\\
&&[M_{\hi\hj},M_{\hk\hl}]=\omega_{\hj\hk}M_{\hi\hl}+\omega_{\hi\hk}M_{\hj\hl}
+\omega_{\hi\hl}M_{\hj\hk}+\omega_{\hj\hl}M_{\hi\hk},\nonumber\\
&&[M_{\bi\bj},Q_{\bk\hk}]=\d_{\bj\bk}Q_{\bi\hk}-\d_{\bi\bk}Q_{\bj\hk},\nonumber\\
&&[M_{\hi\hj},Q_{\bk\hk}]=\omega_{\hj\hk}Q_{\bk\hi}+\omega_{\hi\hk}Q_{\bk\hj},\nonumber\\
&&[Q_{\bi\hi},Q_{\bj\hj}]=k(\omega_{\hi\hj}M_{\bi\bj}+\d_{\bi\bj}M_{\hi\hj}),
\end{eqnarray}
Similarly, we denote
the fermionic generators and the antisymmetric tensor of $OSp(N_2|2N_3)$ as
\e
Q_\pa=Q_{\bi\pri}\quad {\rm and} \quad
\omega_{\pa\pb}=\omega_{\bi\pri,\bj\pj}=\d_{\bi\bj}\omega_{\pri\pj}.
\ee
where $\pri=1,\cdots, 2N_3$ is an $Sp(2N_3)$ fundamental index, which is \emph{independent} of $\hi$ in the sense that
\e\label{cmt}
[M_{\pri\pj}, Q_{\bi\hi}]=0\quad {\rm and }\quad [M_{\hi\hj}, Q_{\bi\pri}]=0.
\ee
Eqs. (\ref{cmt}) are explicit examples of (\ref{trivialcm}). The super Lie algebra $OSp(N_2|2N_3)$ has similar expressions as that of (\ref{OSpfus}).

To construct the corresponding $\CN=4$ theory, we can calculate the double graded commutator
\e
[\{Q_a,Q_b\},Q_\pc]=f_{ab\pc}{}^\pd Q_\pd
\ee
and read off the structure constants from the right hand side. Using (\ref{OSpfus}), we obtain
\begin{eqnarray}
[\{Q_{\bi\hat{i}}, Q_{\bj\hat{j}}\},
Q_{\bk\pk}]=k\omega_{\hi\hj}(\d_{\bj\bk}Q_{\bi\pk}-\d_{\bi\bk}Q_{\bj\pk}).
\end{eqnarray}
It is not difficult to read off the structure constants $f_{ab\pc\pd}$:
\begin{equation}\label{str1}
f_{ab\pc\pd}=f_{\bi\hi,\bj\hj,\bk\pk,\bl\pl}=k\omega_{\hi\hj}\omega_{\pk\pl}
(\d_{\bi\bk}\d_{\bj\bl}-\d_{\bi\bl}\d_{\bj\bk}).
\end{equation}
Similarly, one can calculate $f_{abcd}$ by using $[\{Q_a,Q_b\},Q_c]=f_{abc}{}^d Q_d$.
A short calculation gives
\begin{eqnarray}\label{str2}
f_{abcd}=f_{\bi\hat{i},\bj\hat{j},\bk\hat{k},\bl\hat{l}}
=k[(\delta_{\bi\bk}\delta_{\bj\bl}-\delta_{\bi\bl}\delta_{\bj\bk})
\omega_{\hat{i}\hat{j}}\omega_{\hat{k}\hat{l}}
-\delta_{\bi\bj}\delta_{\bk\bl}(\omega_{\hat{i}\hat{k}}
\omega_{\hat{j}\hat{l}}+\omega_{\hat{i}\hat{l}}
\omega_{\hat{j}\hat{k}})].
\end{eqnarray}
And $f_{\pa\pb\pc\pd}$ have a similar expression:
\begin{equation}\label{str3}
f_{\pa\pb\pc\pd}=f_{\bi\pri,\bj\pj,\bk\pk,\bl\pl}
=k[(\delta_{\bi\bk}\delta_{\bj\bl}-\delta_{\bi\bl}\delta_{\bj\bk})
\omega_{\pri\pj}\omega_{\pk\pl}
-\delta_{\bi\bj}\delta_{\bk\bl}(\omega_{\pri\pk}
\omega_{\pj\pl}+\omega_{\pri\pl} \omega_{\pj\pk})].
\end{equation}

Alternatively, one can read off $k_{uv}$ and $\t^u_{ab}$ from
(\ref{OSpfus}) by comparing (\ref{OSpfus}) with (\ref{slie2}) as well as (\ref{dcanti}).
For instance,
\begin{eqnarray}\label{ospnp}
(\t_{\bm\bn})_{\bi\hi,\bj\hj}&=&\omega_{\hi\hj}(\d_{\bm\bi}\d_{\bn\bj}-\d_{\bm\bj}\d_{\bn\bi}),\\
k^{\bm\bn,\brp\bq}&=&\frac{k}{4}(\d^{\bm\brp}\d^{\bn\bq}-\d^{\bm\bq}\d^{\bn\brp}).
\end{eqnarray}
Similarly, we have
\begin{eqnarray}\label{ospp}
(\t_{\brp\bq})_{\bk\pk,\bl\pl}&=&\omega_{\pk\pl}(\d_{\brp\bk}\d_{\bq\bl}-\d_{\brp\bl}\d_{\bq\bk}).
\end{eqnarray}
Combining Eqs. (\ref{ospnp})$-$(\ref{ospp}) gives (\ref{str1}):
\begin{equation}\label{str1a}
f_{ab\pc\pd}=k_{gh}\t^g_{ab}\t^h_{\pc\pd}=k^{\bm\bn,\brp\bq}
(\t_{\bm\bn})_{\bi\hi,\bj\hj}(\t_{\brp\bq})_{\bk\pk,\bl\pl}
=f_{\bi\hi,\bj\hj,\bk\pk,\bl\pl}=k\omega_{\hi\hj}\omega_{\pk\pl}
(\d_{\bi\bk}\d_{\bj\bl}-\d_{\bi\bl}\d_{\bj\bk}).
\end{equation}
In this way, one can also calculate
$f_{abcd}=k_{uv}\t^u_{ab}\t^v_{cd}$ and
$f_{\pa\pb\pc\pd}=k_{\pu\pv}\t^\pu_{\pa\pb}\t^\pv_{\pc\pd}$; they
are the same as (\ref{str2}) and (\ref{str3}), respectively.

Eqs. (\ref{str1})$-$(\ref{str3}) satisfy the symmetry
conditions (\ref{symfs}), the reality conditions (\ref{rltcndtn})
and the FIs (\ref{FI4}). Eqs. (\ref{str2}) and (\ref{str3}) also
satisfy the constraint equations (\ref{Constr3}). 
Substituting Eqs. (\ref{str1})$-$(\ref{str3}) into
(\ref{LN4}) and (\ref{SUSY4}) gives the $\CN=4$ CSM theory with
gauge group $Sp(2N_1)\times SO(N_2)\times Sp(2N_3)$, which was
first constructed in Ref. \cite{HosomichiJD} by using an ordinary
Lie algebra approach.

\subsection{Fusing $OSp(N_2|2N_1)$ and $OSp(N_2|2N_3)$ into $OSp(N_2|2(N_1+N_3))$}\label{secfsosp1}
In this section we investigate the two superalgebras
$OSp(N_2|2N_1)$ and $OSp(N_2|2N_3)$ further. We demonstrate that
they can be ``fused" into a single closed superalgebra by using
two distinct methods. Finally we prove that the ``fused"
superalgebra is nothing but $OSp(N_2|2(N_1+N_3))$.

Here the essential observation is that the
anti-commutator of $Q_a$ and $Q_\pb$ cannot vanish, i.e.
\e\label{qqprime}
\{Q_{a},Q_{\pb}\}=\{Q_{\bi\hi},Q_{\bj\pj}\}\neq0
\ee
\emph{provided that the $Q_aQ_\pb Q_\pc$ ($Q_{\bi\hi}Q_{\bj\pj}Q_{\bk\pk}$) Jacobi identity is obeyed.}
Actually,
if $\{Q_{\bi\hi},Q_{\bj\pj}\}=0$, then the
$Q_{\bi\hi}Q_{\bj\pj}Q_{\bk\pk}$ Jacobi identity
\e
[\{Q_{\bi\hi},Q_{\bj\pj}\},Q_{\bk\pk}]+[\{Q_{\bi\hi},Q_{\bk\pk}\},Q_{\bj\pj}]+
[\{Q_{\bk\pk},Q_{\bj\pj}\},Q_{\bi\hi}]=0
\ee
 implies that
\begin{equation}[M_{\bj\bk},Q_{\bi\hi}]=0,\end{equation}
which is contradictory with the third equation of (\ref{OSpfus}).
So (\ref{qqprime}) must hold. According to the fundamental idea of
supersymmetry, the anti-commutator of two fermionic generators
must be a linear combination of bosonic generators. On the other
hand, since $\hi$ and $\pj$ are independent indices, it is natural
to define
\begin{eqnarray}\label{newcom}
&&\{Q_{\bi\hi},Q_{\bj\pj}\}=k\d_{\bi\bj}M_{\hi\pj},\quad
[M_{\hi\hj},Q_{\bk\pl}]=[M_{\pri\pj},Q_{\bk\hl}]=0,\nonumber\\
&&[M_{\hi\pri},Q_{\bj\pj}]=\omega_{\pri\pj}Q_{\bj\hi},\quad
[M_{\hi\pri},Q_{\bj\hj}]=\omega_{\hi\hj}Q_{\bj\pri}.
\end{eqnarray}
In the first equation, we have introduced a set of new bosonic
generators $M_{\hi\pj}$. So the first equation of (\ref{newcom})
is an explicit example of (\ref{ntrvanticm0}). Using
(\ref{newcom}), it is easy to verify that the
$Q_{\bi\hi}Q_{\bj\pj}Q_{\bk\pk}$ Jacobi identity and the
$Q_{\bi\hi}Q_{\bj\hj}Q_{\bk\pk}$ Jacobi identity are obeyed. One
can also define all other possible commutators involving
$M_{\hi\pj}$, i.e. $[M_{\hi\pri}, M_{\hj\pj}]$,
$[M_{\hi\hj},M_{\hk\pk}]$ and $[M_{\pri\pj},M_{\hk\pk}]$, by
requiring that the corresponding Jacobi identities are obeyed. For
example, consider the $M_{\hi\pri}Q_{\hj\pj}Q_{\hk\pk}$ Jacobi
identity
\begin{equation}
[M_{\hi\pri},\{Q_{\bj\hj},Q_{\bk\pk}\}]-\{Q_{\bj\hj},[M_{\hi\pri},Q_{\bk\pk}]\}
-\{Q_{\bk\pk},[M_{\hi\pri},Q_{\bj\hj}]\}=0.
\end{equation}
A short calculation gives
\begin{equation}\label{pmpm}
[M_{\hi\pri},M_{\hj\pk}]=\omega_{\pri\pk}M_{\hi\hj}+\omega_{\hi\hj}M_{\pri\pk}.
\end{equation}
It can be seen that the commutator of two new generators gives
rise an $Sp(2N_1)$ generator $M_{\hi\hj}$ of $G$ and an $Sp(2N_3)$
generator $M_{\pri\pk}$ of $G^\prime$. The structure constants of
the commutator (\ref{pmpm}) furnish a fundamental representation
of $M_{\hi\hj}$ and a fundamental representation of $M_{\pri\pk}$.
Similarly, the commutator $[M_{\hi\hj},M_{\hk\pk}]$, determined by
the $M_{\hi\hj}Q_{\bk\hk}Q_{\bj\pk}$ Jacobi identity, is given by
\begin{equation}\label{unpmnew}
[M_{\hi\hj},M_{\hk\pk}]=\omega_{\hj\hk}M_{\hi\pk}+\omega_{\hi\hk}M_{\hj\pk}.
\end{equation}
The structure constants of the above commutator furnish a
fundamental representation of $M_{\hi\hj}$. Finally the
$M_{\pri\pj}Q_{\bk\hk}Q_{\bj\pk}$ Jacobi identity gives
\begin{equation}\label{pmnew}
[M_{\pri\pj},M_{\hk\pk}]=\omega_{\pj\pk}M_{\hk\pri}+\omega_{\pri\pk}M_{\hk\pj}.
\end{equation}
It can be seen that $M_{\hk\pk}$ provide a fundamental representation of $M_{\pri\pj}$.

It is not difficult (though a little tedious) to verify that
\emph{every} Jacobi identity of the ``total" superalgebra
consisting of $M_{\hi\pj}$ \emph{and} all generators of
$OSp(N_2|2N_1)$ \emph{and} $OSp(N_2|2N_3)$ is satisfied. So the
five graded commutators in (\ref{newcom}) must be the correct
ones, and the new superalgebra ``fused"
by $OSp(N_2|2N_1)$ and $OSp(N_2|2N_3)$ is \emph{closed}. 
The commutation relations of the ``fused" superalgebra include
(\ref{newcom}), (\ref{pmpm}), (\ref{unpmnew}), (\ref{pmnew}), and
the commutation relations of $OSp(N_2|2N_1)$ and $OSp(N_2|2N_3)$.
These commutation relations suggest that the ``fused" superalgebra
is \emph{simple}, i.e. it has no proper invariant
sub-superalgebra.

In this way, we have ``fused" the two orthosymplectic
superalgebras $OSp(N_2|2N_1)$ and $OSp(N_2|2N_3)$ by solving the
important Jacobi identities. We now want to provide an alternative
way to construct the ``fused" superalgebra.  The main idea is the
following: Using oscillators to realize $OSp(N_2|2N_1)$ and
$OSp(N_2|2N_3)$ first, then all commutation relations for fusing
$OSp(N_2|2N_1)$ and $OSp(N_2|2N_3)$ can be determined
straightforwardly by using oscillator algebras.

To realize $OSp(N_2|2N_1)$, we introduce a set of bosonic
oscillators and a set of fermionic oscillators as follows
\e\label{oscillators} &&[b_{\bi},
b^{\bj\dag}]=\d_{\bi}^{\bj},\quad [b_{\bi}, b_{\bj}]=[b^{\bi\dag},
b^{\bj\dag}]=0;\\\label{oscillators2} &&\{a_{\hi},
a^{\hj\dag}\}=\d_{\hi}^{\hj},\quad \{a_{\hi},
a_{\hj}\}=\{a^{\hi\dag}, a^{\hj\dag}\}=0. \ee Here
$\bi=1,\ldots,N_2$ and $\hi=1,\ldots, 2N_1$. We use the invariant
tensors $\d_{\bi\bj}$ and $\omega_{\hi\hj}$ to lower indices; for
instance, \e b^{\dag}_{\bi}\equiv\d_{\bi\bj}b^{\bj\dag},\quad {\rm
and }\quad a^\dag_{\hi}\equiv\omega_{\hi\hj}a^{\hj\dag}. \ee The
generators of $OSp(N_2|2N_1)$ can be constructed as follows
\e\label{osciOSp1}
Q_{\bi\hi}=\sqrt{-k}(a_{\hi}b^\dag_{\bi}+a^\dag_{\hi}b_{\bi}),\quad
M_{\bi\bj}=b^\dag_{\bi}b_{\bj}-b^\dag_{\bj}b_{\bi},\quad
M_{\hi\hj}=-(a^\dag_{\hi}a_{\hj}+a^\dag_{\hj}a_{\hi}). \ee It is
straightforward to verify that (\ref{osciOSp1}) satisfy the
commutation relations of $OSp(N_2|2N_1)$ (\ref{OSpfus}).
Similarly, the generators of $OSp(N_2|2N_3)$ can be constructed as
follows \e\label{osciOSp2}
Q_{\bi\pri}=\sqrt{-k}(c_{\pri}b^\dag_{\bi}+c^\dag_{\pri}b_{\bi}),\quad
M_{\bi\bj}=b^\dag_{\bi}b_{\bj}-b^\dag_{\bj}b_{\bi},\quad
M_{\pri\pj}=-(c^\dag_{\pri}c_{\pj}+c^\dag_{\pj}c_{\pri}), \ee
where $b_{\bi}$ and $b^{\dag}_{\bi}$ are the \emph{same} as that
of (\ref{oscillators}); $c_{\pri}$ and
$c^\dag_{\pri}\equiv\omega_{\pri\pj}c^{\pj\dag}$
$(\pri=1,\ldots,2N_3)$ are a third independent set of oscillators,
satisfying \e
&&\{c_{\pri}, c^{\pj\dag}\}=\d_{\pri}^{\pj},\quad \{c_{\pri}, c_{\pj}\}=\{c^{\pri\dag}, c^{\pj\dag}\}=0.
\ee
With (\ref{osciOSp1}) and (\ref{osciOSp2}), the anticommutator of $Q_{\bi\hi}$ and $Q_{\bj\pj}$ is given by
\e\label{mixcmmt}
\{Q_{a},Q_{\pb}\}=\{Q_{\bi\hi}, Q_{\bj\pj}\}=k\d_{\bi\bj}(-c^\dag_{\pj}a_{\hi}+c_{\pj}a^\dag_{\hi}).
\ee
Comparing it with the first equation of (\ref{newcom}), we are led to define the set of new bosonic generators $M_{\hi\pj}$ as
\e\label{osciOSp30}
M_{\hi\pj}=-c^\dag_{\pj}a_{\hi}+c_{\pj}a^\dag_{\hi}.
\ee
Substituting the oscillator realizations (\ref{osciOSp1}), (\ref{osciOSp2}), and (\ref{osciOSp30}) into the commutation relations (\ref{newcom}), (\ref{pmpm}), (\ref{unpmnew}), and (\ref{pmnew}), we find that they are exactly obeyed.
Namely, all the generators constructed in terms of oscillators obey exactly the same commutation relations as before. It is therefore unnecessarily to verify the Jacobi identities. In this way, we have constructed the closed superalgebra ``fused" by $OSp(N_2|2N_1)$ and $OSp(N_2|2N_3)$ in terms of \emph{three} independent sets of oscillators. The advantage of the oscillator-realization approach is that it shows explicitly that it is unavoidable to introduce the set of new bosonic generators $M_{\hi\pj}$ (see (\ref{mixcmmt}) and (\ref{osciOSp30})). Also, using oscillators one can construct the commutation relations (\ref{newcom}), (\ref{pmpm}), (\ref{unpmnew}), and (\ref{pmnew}) without any guessing work.

Let us summarize the bosonic subalgebra of the ``fused" superalgebra as follows:
\begin{eqnarray}\label{bosonicpart}
&&[M_{\bar i\bar j},M_{\bar k\bar
l}]=\d_{\bj\bk}M_{\bi\bl}-\d_{\bi\bk}M_{\bj\bl}+\d_{\bi\bl}M_{\bj\bk}
-\d_{\bj\bl}M_{\bi\bk},\nonumber\\
&&[M_{\hi\hj},M_{\hk\hl}]=\omega_{\hj\hk}M_{\hi\hl}+\omega_{\hi\hk}M_{\hj\hl}
+\omega_{\hi\hl}M_{\hj\hk}+\omega_{\hj\hl}M_{\hi\hk},
\nonumber\\
&&[M_{\pri\pj},M_{\pk\pl}]=\omega_{\pj\pk}M_{\pri\pl}+\omega_{\pri\pk}M_{\pj\pl}
+\omega_{\pri\pl}M_{\pj\pk}+\omega_{\pj\pl}M_{\pri\pk},
\nonumber\\&&
[M_{\hi\pri},M_{\hj\pk}]=\omega_{\pri\pk}M_{\hi\hj}+\omega_{\hi\hj}M_{\pri\pk},\nonumber\\
&&
[M_{\hi\hj},M_{\hk\pk}]=\omega_{\hj\hk}M_{\hi\pk}+\omega_{\hi\hk}M_{\hj\pk},\nonumber\\
&&
[M_{\pri\pj},M_{\hk\pk}]=\omega_{\pj\pk}M_{\hk\pri}+\omega_{\pri\pk}M_{\hk\pj}.
\end{eqnarray}
The other commutators vanish. The first 3 lines are the Lie algebras of $SO(N_2)$, $Sp(2N_1)$, and $Sp(2N_3)$, respectively; the last 3 lines are the commutators involving the set of new generators $M_{\hi\pri}$ (see (\ref{pmpm}), (\ref{unpmnew}), and (\ref{pmnew})).
So the bosonic part of the `fused' superalgebra consists of \emph{four} sets of generators
\e\label{full}
M^U=(M_{\hi\hj},M_{\bi\bj},M_{\pri\pj},M_{\hi\pj}),
\ee
in which we have selected only the first \emph{three} sets of generators , namely
\e\label{unfull}
M^m=(M_{\hi\hj},M_{\bi\bj},M_{\pri\pj}),
\ee
to construct the $\CN=4$ CSM theory with
gauge group $Sp(2N_1)\times SO(N_2)\times Sp(2N_3)$. Notice that (\ref{unfull}) is an example of (\ref{drcsm1}).

It can be seen that the Lie algebra of $SO(N_2)$ (the first line of (\ref{bosonicpart})), the common bosonic part of $OSp(N_2|2N_1)$ and $OSp(N_2|2N_3)$, is an invariant subalgebra of (\ref{bosonicpart}). We now prove that the last five lines of (\ref{bosonicpart}) are actually the commutation relations of $Sp(2(N_1+N_3))$. Let us begin by considering the simplest case, i.e. $N_1=N_3=1$. It is convenient to define
\e
N^a=i\s^{a\dag\pri\hi}M_{\hi\pri},\quad {\rm and }\quad M^{ab}=\frac{1}{2}(\bar\s^{ab\pri\pj}M_{\pri\pj}+\s^{ab\hi\hj}M_{\hi\hj}).
\ee
Here $\s^a$, $\s^{ab}$ and $\bar\s^{ab}$ are defined as
\e\label{paulim}
&&\sigma^a=(\sigma^1,\sigma^2,\sigma^3,
i\mathbb{I}),\quad
\sigma^{a\dag}=(\sigma^1,\sigma^2,\sigma^3,
-i\mathbb{I}),\\
&&\s^{ab}=\frac{1}{4}(\s^a\s^{b\dag}-\s^b\s^{a\dag}),\quad
\bar\s^{ab}=\frac{1}{4}(\s^{a\dag}\s^b
-\s^{b\dag}\s^a),\label{su2su2}
\ee%
where $\s^{i}$ ($i=1,\ldots, 3$) are Pauli matrices and
$\mathbb{I}$ is the $2\times2$ unit matrix. After some algebraic
steps, we convert the last five lines of (\ref{bosonicpart}) into
the form \e\label{so5}
&&[N^a,N^b]=4M^{ab},\nonumber\\
&&[M^{ab},M^{cd}]=\d^{bc}M^{ad}-\d^{ac}M^{bd}-\d^{bd}M^{ac}+\d^{ad}M^{bc},\nonumber\\
&&[M^{ab},N^c]=\d^{bc}N^a-\d^{ac}N^b.
\ee
The second line is the familiar Lie algebra of $SO(4)$. However, after combining the first line and the last line, the algebra turns out to be the Lie algebra of $SO(5)$. To see this, define
\e
M^{a5}=-M^{5a}=\frac{i}{2}N^a\quad {\rm and}\quad M^{55}=0.
\ee
Now (\ref{so5}) can be recast into
\e\label{so52}
[M^{ij},M^{kl}]=\d^{jk}M^{il}-\d^{ik}M^{jl}-\d^{jl}M^{ik}+\d^{il}M^{jk},
\ee
where $i=1,\ldots,5$. Eq. (\ref{so52}) is nothing but the Lie algebra of $SO(5)$.

Recall that the Lie algebra of $SO(5)$ is isomorphic to that of $Sp(4)$. So in the special case of $N_1=N_3=1$, the last five lines of (\ref{bosonicpart}) are indeed the commutation relations of $Sp(2(N_1+N_3))=Sp(4)$, hence (\ref{bosonicpart}) is nothing but the Lie algebra of $SO(N_2)\times Sp(4)$. On the other hand, in Section \ref{secexample}, we have observed that the ``fused" superalgebra is simple. A superalgebra whose bosonic part is the Lie algebra of $SO(N_2)\times Sp(4)$ must be $OSp(N_2|4)$. This special case inspires us to guess that for general $N_1$ and $N_3$, the closed superalgebra ``fused" by $OSp(N_2|2N_1)$ and $OSp(N_2|2N_3)$ is nothing but $OSp(N_2|2(N_1+N_3))$. To prove it, we combine the fermionic generators $Q_{\bi\hi}$ and $Q_{\bi\pri}$ as follows
\e\label{cmb2osci}
Q_{\bi I}=\begin{pmatrix} Q_{\bi\hi}\\Q_{\bi\pri}\end{pmatrix}=Q_{\bi\hi}\d_{1\a}+Q_{\bi\pri}\d_{2\a},
\ee
where $I=1,\ldots, 2(N_1+N_3)$ is a collective index; $\d_{1\a}=(1,0)^{\rm T}$ and $\d_{2\a}=(0,1)^{\rm T}$ are ``spin up" spinor and ``spin down" spinor, respectively (they are \emph{not} spacetime spinors).

In the oscillator realization, Eq. (\ref{cmb2osci}) takes the form
\e
Q_{\bi I}=\sqrt{-k}(A^\dag_Ib_{\bi}+A_Ib^\dag_{\bi}),
\ee
where we have combined the two independent sets \emph{fermionic} oscillators as one set:
\e\label{twoosci}
&&A_I=\begin{pmatrix} a_{\hi}\\c_{\pri}\end{pmatrix},\quad A^{I\dag}=\begin{pmatrix} a^{\hi\dag}\\c^{\pri\dag}\end{pmatrix},\quad
\{A_{I}, A^{J\dag}\}=\d_{I}^{J},\quad \{A_{I}, A_{J}\}=\{A^{I\dag}, A^{J\dag}\}=0.
\nonumber\\&&\omega_{IJ}=\begin{pmatrix}\omega_{\hi\hj} & 0\\0 &\omega_{\pri\pj} \end{pmatrix},\quad A_I^\dag\equiv \omega_{IJ}A^{J\dag}.
\ee
With the above collective notation, all anti-commutators of the ``fused" superalgebra can be summed up in the single one,
\e\label{anticm}
\{Q_{\bi I},Q_{\bj J}\}=k(\d_{\bi\bj}M_{IJ}+\omega_{IJ}M_{\bi\bj}),
\ee
where we have defined
\e\label{ospcmtt}
M_{IJ}&=&M_{\hi\hj}\d_{1\a}\d_{1\b}+M_{\hi\pj}\d_{1\a}\d_{2\b}+ M_{\hj\pri}\d_{2\a}\d_{1\b}+M_{\pri\pj}\d_{2\a}\d_{2\b},\\
\omega_{IJ}&=&\omega_{\hi\hj}\d_{1\a}\d_{1\b}+\omega_{\pri\pj}\d_{2\a}\d_{2\b}\label{ospomega}.
\ee
Eq. (\ref{ospomega}) is just the component formalism of $\omega_{IJ}$ defined in (\ref{twoosci}). On the other hand, all commutators between the bosonic generators and the fermionic generators are compacted into two commutators:
\e\label{cmm}
&&[M_{IJ},Q_{\bk K}]=\omega_{JK}Q_{\bk I}+\omega_{IK}Q_{\bk J},\nonumber\\
&&[M_{\bi\bj},Q_{\bk K}]=\d_{\bj\bk}Q_{\bi K}-\d_{\bi\bk}Q_{\bj K}.
\ee
With (\ref{ospcmtt}) and (\ref{ospomega}), after a lengthy algebra, the last five lines of (\ref{bosonicpart}) can be recast into
\e\label{sptt}
[M_{IJ},M_{KL}]=\omega_{JK}M_{IL}+\omega_{IK}M_{JL}+\omega_{JL}M_{IK}+\omega_{IL}M_{JK},
\ee
which is nothing but the Lie algebra of $Sp(2(N_1+N_3))$. Together with the algebra of $SO(N_2)$ (see the first commutator of (\ref{OSpfus})), (\ref{anticm}), (\ref{cmm}), and (\ref{sptt}) are precisely the commutation relations of $OSp(N_2|2(N_1+N_3))$. This completes the proof. Notice that the bosonic subalgebra of $OSp(N_2|2(N_1+N_3))$ is the
Lie algebra of $SO(N_2)\times Sp(2(N_1+N_3))$, while the Lie algebra of the gauge group is the Lie algebra of $Sp(2N_1)\times SO(N_2)\times Sp(2N_3)$. Namely, we used only the bosonic part of $OSp(N_2|2N_1)$ \emph{and} the bosonic part of $OSp(N_2|2N_3)$, \emph{not} the bosonic part of $OSp(N_2|2(N_1+N_3))$, to construct the physical theory. The difference is precisely the difference between (\ref{full}) and (\ref{unfull}).


Notice that two or more theories with different gauge groups may be associated the \emph{same} ``fused" superalgebra. For instance, if we select the bosonic parts of $OSp(N_2|2(N_1-\lambda))$ and $OSp(N_2|2(N_3+\lambda))$ ($\lambda=0,\ldots,N_1-1$), sharing the common factor $SO(N_2)$, as the Lie algebra
of the gauge group of the $\CN=4$ theory, we will obtain different theories (with different gauge groups) as $\lambda$ runs from $0$ to $(N_1-1)$. However, the corresponding ``fused" superalgebras are independent of $\lambda$, since
\e\nonumber
&&\bigg(OSp(N_2|2(N_1-\lambda))\quad {\rm fusing }\quad OSp(N_2|2(N_3+\lambda))\bigg)\\&=&OSp(N_2|2(N_1-\lambda+N_3+\lambda))=OSp(N_2|2(N_1+N_3)).
\ee

\subsection{Fusing $OSp(N_2|2N_1)$ and $OSp(N_4|2N_1)$ into $OSp(N_2+N_4|2N_1)$}\label{secfsosp2}

In this section, we choose the superalgebras $G$ and $G^\prime$ as $OSp(N_2|2N_1)$ and $OSp(N_4|2N_1)$, respectively. The common simple factor of their bosonic parts is the Lie algebra of $Sp(2N_1)$. The commutation relations of $OSp(N_2|2N_1)$ are given by (\ref{OSpfus}). We denote the fermionic generators of $OSp(N_2|2N_1)$ and $OSp(N_4|2N_1)$ as
\e
Q_a=Q_{\bi\hi} \quad{\rm and}\quad Q_\pa=Q_{\pri\hi},
\ee
respectively.  Here $\bi=1,\cdots,N_2$ is an $SO(N_2)$ fundamental index;
$\hi=1,\cdots, 2N_1$ is an $Sp(2N_1)$ fundamental index; $\pri=1,\cdots,N_4$ is an $SO(N_4)$ fundamental index. The structure constants of the 3-algebra $f_{ab\pc\pd}$ are identified with the structure constants of the double graded commutator on $G$ and $G^\prime$: $[\{Q_a, Q_b\},Q_{\pc}]=f_{ab\pc}{}^{\pd}Q_{\pd}$.
A short calculation gives
\e
f_{ab\pc\pd}=f_{\bi\hi,\bj\hj,\pk\hk,\pl\hl}=-k\d_{\bi\bj}\d_{\pk\pl}
(\omega_{\hi\hk}\omega_{\hj\hl}+\omega_{\hj\hk}\omega_{\hi\hl}).
\ee
Also, $f_{abcd}$ are given by (\ref{str2}) and $f_{\pa\pb\pc\pd}$ have a similar expression as that of (\ref{str2}). Substituting these structure constants into the action (\ref{LN4}) and the supersymmetry transformations (\ref{SUSY4}) gives the $\CN=4$ CSM theory with gauge group $SO(N_2)\times Sp(2N_1)\times SO(N_4)$. This theory was first constructed in Ref.
\cite{HosomichiJD}, using an ordinary Lie algebra approach.

To fuse the two superalgebras $G$ and $G^\prime$, we first use oscillators to realize them. The oscillator realization of $OSp(N_2|2N_1)$ is given by (\ref{oscillators})$-$(\ref{osciOSp1}); and $OSp(N_4|2N_1)$ has a similar realization:
\e\label{osciOSp3}
Q_{\pri\hi}=\sqrt{-k}(a_{\hi}d^\dag_{\pri}+a^\dag_{\hi}d_{\pri}),\quad M_{\pri\pj}=d^\dag_{\pri}d_{\pj}-d^\dag_{\pj}d_{\pri},\quad M_{\hi\hj}=-(a^\dag_{\hi}a_{\hj}+a^\dag_{\hj}a_{\hi}).
\ee
where $a_{\hi}$ and $a^{\dag}_{\hi}$ are the \emph{same} as that of (\ref{oscillators2}); $d_{\pri}$ and $d^\dag_{\pri}\equiv\d_{\pri\pj}d^{\pj\dag}$ $(\pri=1,\ldots,N_4)$ are a third independent set of oscillators, satisfying
\e
[d_{\pri}, d^{\pj\dag}]=\d_{\pri}^{\pj},\quad [d_{\pri}, d_{\pj}]=[d^{\pri\dag}, d^{\pj\dag}]=0.
\ee
With a little oscillator algebra, we obtain
\e\label{nicecmtt}
\{Q_{a},Q_{\pb}\}=\{Q_{\bi\hi},Q_{\pj\hj}\}= k\omega_{\hi\hj}M_{\bi\pj},\quad M_{\bi\pj}=b^\dag_{\bi}d_{\pj}-b_{\bi}d^
\dag_{\pj}.
\ee
This provides another example for the assertion in Section \ref{Introduction} that if the bosonic parts of two superalgebras $G$ and $G^\prime$ share one simple factor, then the anticommutator between their fermionic generators cannot vanish, i.e. $\{Q_{a},Q_{\pb}\}\neq0$. In order to use the same technique as the previous section, we define
\e\label{2fmges}
Q_{I\hi}=\begin{pmatrix} Q_{\bi\hi}\\Q_{\pri\hi}\end{pmatrix}=Q_{\bi\hi}\d_{1\a}+Q_{\pri\hi}\d_{2\a},
\ee
where $I=1,\ldots, (N_2+N_4)$ is a collective index; $\d_{1\a}=(1,0)^{\rm T}$ and $\d_{2\a}=(0,1)^{\rm T}$ are ``spin up" spinor and ``spin down" spinor respectively (they are \emph{not} spacetime spinors). In the oscillator construction, Eq. (\ref{2fmges}) is given by
\e
Q_{I\hi}=\sqrt{-k}(B^\dag_Ia_{\hi}+B_Ia^\dag_{\hi})
\ee
where we have combined the two independent sets \emph{bosonic} oscillators as one set:
\e\label{twoosci2}
&&B_I=\begin{pmatrix} b_{\bi}\\d_{\pri}\end{pmatrix},\quad B^{I\dag}=\begin{pmatrix} b^{\bi\dag}\\d^{\pri\dag}\end{pmatrix},\quad
[B_{I}, B^{J\dag}]=\d_{I}^{J},\quad [B_{I}, B_{J}]=[B^{I\dag}, B^{J\dag}]=0.
\nonumber\\&&\d_{IJ}=\begin{pmatrix}\d_{\bi\bj} & 0\\0 &\d_{\pri\pj} \end{pmatrix},\quad B_I^\dag\equiv \d_{IJ}B^{J\dag}.
\ee
With the above compact notation, we obtain
\e\label{anticm2}
\{Q_{I\hi},Q_{J\hj}\}=k(\d_{IJ}M_{\hi\hj}+\omega_{\hi\hj}M_{IJ}),\quad
[M_{\hi\hj},Q_{K\hk}]=\omega_{\hj\hk}Q_{K\hi}+\omega_{\hi\hk}Q_{K\hj},
\ee
where we have defined
\e\label{ospcmtt2}
&&M_{IJ}=M_{\bi\bj}\d_{1\a}\d_{1\b}+M_{\bi\pj}\d_{1\a}\d_{2\b}- M_{\bj\pri}\d_{2\a}\d_{1\b}+M_{\pri\pj}\d_{2\a}\d_{2\b},\\
&&\d_{IJ}=\d_{\bi\bj}\d_{1\a}\d_{1\b}+\d_{\pri\pj}\d_{2\a}\d_{2\b}\label{ospcmtt2q}.
\ee
Eq. (\ref{ospcmtt2q}) is the component formalism of $\d_{IJ}$ defined in (\ref{twoosci2}). On the other hand, by either requiring that the $Q_{I\hi}Q_{J\hj}Q_{K\hk}$ Jacobi identity is obeyed or using straightforward oscillator algebra, we can obtain
\e\label{cmm2}
[M_{IJ},Q_{K\hk}]=\d_{JK}Q_{I\hk}-\d_{IK}Q_{J\hk}.
\ee
Similarly, by either requiring that the $Q_{I\hi}Q_{J\hj}M_{KL}$ Jacobi identity is obeyed or using a rather lengthy oscillator algebra, we can derive the commutator
\e\label{sptt2}
[M_{IJ},M_{KL}]=\d_{JK}M_{IL}-\d_{IK}M_{JL}-\d_{JL}M_{IK}+\d_{IL}M_{JK},
\ee
which is just the Lie algebra of $SO(N_2+N_4)$. Together with the algebra of $Sp(N_1)$ (the second line of (\ref{OSpfus})), (\ref{anticm2}), (\ref{cmm2}), and (\ref{sptt2}) are precisely the commutation relations of $OSp(N_2+N_4|2N_1)$. Although we have completed the ``fusion" of $G$ and $G^\prime$ in terms of their oscillator realizations, the commutation relations (\ref{nicecmtt}), (\ref{anticm2}), (\ref{cmm2}), and (\ref{sptt2}) must hold in general.

\section{Fusing $U(N_1|N_2)$ and $U(N_2|N_3)$ into $U(N_1+N_3|N_2)$}\label{secfumun}
In this section, we select the superalgebras $G$ and $G^\prime$ as $U(N_1|N_2)$ and $U(N_2|N_3)$, respectively. Namely, the common part of their bosonic subalgebras is the Lie algebra of $U(N_2)$, which is \emph{not} simple due to the fact that $U(N_2)=SU(N_2)\times U(1)$. The commutation relations of $U(M|N)$ are given by Appendix \ref{cmrunum}. We denote the fermionic generators of $U(N_2|N_3)$ as $Q_{\bu}{}^\pri$ and $\bar Q_\pri{}^{\bu}$, where the subscript index $\bu=1,\ldots, N_2$ is a fundamental index of $U(N_2)$ and the superscript index $\pri=1,\ldots, N_3$ is an anti-fundamental indices of $U(N_3)$. Similarly, we denote the fermionic generators of $U(N_1|N_2)$ as $Q_{\bu}{}^{\hi}$ and $\bar Q_{\hi}{}^{\bu}$, with $\hi=1,\ldots, N_1$.

We first use $U(N_1|N_2)$ and $U(N_2|N_3)$ to construct the 3-algebra in the $\CN=4$ theory. It is useful to define
\begin{equation}\label{spnpd}
Q_a=\begin{pmatrix} \bar Q_{\hi}{}^{\bu}\\-Q_{\bu}{}^{\hi}\end{pmatrix}=\bar Q_{\hi}{}^{\bu}\d_{1\lambda}-Q_{\bu}{}^{\hi}\d_{2\lambda}, \quad Q_\pa=\begin{pmatrix} \bar Q_\pri{}^{\bu}\\-Q_{\bu}{}^\pri\end{pmatrix}=\bar Q_\pri{}^{\bu}\d_{1\a}-Q_{\bu}{}^\pri\d_{2\a},
\end{equation}
where $\d_{1\lambda}=(1,0)^{{\rm T}}$ and $\d_{2\lambda}=(0,1)^{{\rm T}}$ are ``spin up" spinor and ``spin down" spinor, respectively. As usual, the structure constants of the 3-algebra $f_{ab\pc\pd}$ can be read off from the double graded commutator: $[\{Q_a, Q_b\},Q_{\pc}]=f_{ab\pc}{}^{\pd}Q_{\pd}$.
A short calculation gives
\e\nonumber
f_{ab\pc\pd}&=&-k(\d_{\hi}{}^{\hj}\d_{\bv}{}^{\bw}\d_{\pk}{}^{\pl}\d_{\bt}{}^{\bu}
\d_{1\lambda}\d_{2\xi}\d_{1\g}\d_{2\d}+\d_{\hj}{}^{\hi}\d_{\bu}{}^{\bw}\d_{\pk}{}^{\pl}\d_{\bt}{}^{\bv}
\d_{2\lambda}\d_{1\xi}\d_{1\g}\d_{2\d}\\
&&\quad\quad+\d_{\hi}{}^{\hj}\d_{\bw}{}^{\bu}\d_{\pl}{}^{\pk}\d_{\bv}{}^{\bt}
\d_{1\lambda}\d_{2\xi}\d_{2\g}\d_{1\d}+\d_{\hj}{}^{\hi}\d_{\bw}{}^{\bv}\d_{\pl}{}^{\pk}\d_{\bu}{}^{\bt}
\d_{2\lambda}\d_{1\xi}\d_{2\g}\d_{1\d}).
\ee
Similarly, we obtain the structure constants $f_{abcd}$,
\begin{eqnarray}\label{strct86}
f_{abcd}
&=&f_{\bi}{}^{\hi}{}_{\bk}{}^{\hk},{}_{\hl}{}^{\bl}{}_{\hj}{}^{\bj}\delta_{2\lambda}\delta_{1\xi}\delta_{2\rho}\delta_{1\sigma}
+f_{\bi}{}^{\hi}{}_{\bl}{}^{\hl},{}_{\hk}{}^{\bk}{}_{\hj}{}^{\bj}\delta_{2\lambda}\delta_{1\xi}\delta_{1\rho}\delta_{2\sigma}
\nonumber\\&&+f_{\bj}{}^{\hj}{}_{\bk}{}^{\hk},{}_{\hl}{}^{\bl}{}_{\hi}{}^{\bi}\delta_{1\lambda}\delta_{2\xi}\delta_{2\rho}\delta_{1\sigma}
+f_{\bj}{}^{\hj}{}_{\bl}{}^{\hl},{}_{\hk}{}^{\bk}{}_{\hi}{}^{\bi}\delta_{1\lambda}\delta_{2\xi}\delta_{1\rho}\delta_{2\sigma},
\end{eqnarray}
where
\e\label{strumun2}
f_{\bi}{}^{\hi}{}_{\bj}{}^{\hj},{}_{\hk}{}^{\bk}{}_{\hl}{}^{\bl}
\equiv k(\d_{\hk}{}^{\hi}\d_{\hl}{}^{\hj}\d_{\bj}{}^{\bk}\d_{\bi}{}^{\bl}
-\d_{\hl}{}^{\hi}\d_{\hk}{}^{\hj}\d_{\bi}{}^{\bk}\d_{\bj}{}^{\bl}).
\ee
And $f_{\pa\pb\pc\pd}$ have a similar expression as that of (\ref{strct86}). Substituting these structure constants into (\ref{LN4}) and (\ref{SUSY4}) gives the $\CN=4$, $U(N_1)\times U(N_2)\times U(N_3)$ CSM theory. This theory was first derived in Ref. \cite{HosomichiJD}, using an ordinary Lie algebra approach.

To fuse $U(N_1|N_2)$ and $U(N_2|N_3)$, let us first construct them by the following \emph{three} independent sets of oscillators:
\e\label{oscillators20}
&&\{a_{\hi}, a^{\hj\dag}\}=\d_{\hi}{}^{\hj},\quad \{a_{\hi}, a_{\hj}\}
=\{a^{\hi\dag}, a^{\hj\dag}\}=0;\nonumber\\&&[b_{\bu}, b^{\bv\dag}]=\d_{\bu}{}^{\bv},\quad [b_{\bu}, b_{\bv}]=[b^{\bu\dag}, b^{\bv\dag}]=0;
\\
&&\{c_{\pri}, c^{\pj\dag}\}=\d_{\pri}{}^{\pj},\quad \{c_{\pri}, c_{\pj}\}
=\{c^{\pri\dag}, c^{\pj\dag}\}=0.\nonumber
\ee
Here $\hi=1,\ldots, N_1$, $\bu=1,\ldots,N_2$, and $\pri=1,\ldots, N_3$. In terms of the oscillators, the generators of $U(N_1|N_2)$ are given by
\e\label{osciunis1}
Q_{\bu}{}^{\hi}=\sqrt{-k}b_{\bu}a^{\hi\dag},\quad \bar Q_{\hi}{}^{\bu}=\sqrt{-k}a_{\hi}b^{\bu\dag},\quad M_{\hi}{}^{\hj}=a_{\hi}a^{\hj\dag},\quad M_{\bu}{}^{\bv}=-b_{\bu}b^{\bv\dag}.
\ee
A little algebra shows that they indeed obey the commutation relations of $U(N_1|N_2)$. Similarly, the generators of $U(N_2|N_3)$ are given by
\e\label{osciunis2}
Q_{\bu}{}^{\pri}=\sqrt{-k}b_{\bu}c^{\pri\dag},\quad \bar Q_{\pri}{}^{\bu}=\sqrt{-k}c_{\pri}b^{\bu\dag},\quad M_{\pri}{}^{\pj}=a_{\pri}a^{\pj\dag},\quad M_{\bu}{}^{\bv}=-b_{\bu}b^{\bv\dag}.
\ee
Now it is easy to calculate the non-trivial anticommutators between the fermionic generators of $U(N_1|N_2)$ and $U(N_2|N_3)$; they are given by
\e
&&\{Q_{\bu}{}^{\hi}, \bar Q_{\pri}{}^{\bv}\}=k\d_{\bu}{}^{\bv}(c_{\pri}a^{\hi\dag})\equiv k\d_{\bu}{}^{\bv}\bar M_{\pri}{}^{\hi},\nonumber\\
&&\{\bar Q_{\hi}{}^{\bu}, Q_{\bv}{}^{\pri}\}=k\d_{\bv}{}^{\bu}(a_{\hi}c^{\pri\dag})\equiv k\d_{\bv}{}^{\bu} M_{\hi}{}^{\pri}.
\ee
Combining them with (\ref{spnpd}), we have
\e\label{qaqpb3}
\{Q_a,Q_\pb\}=-k(\d_{\bv}{}^{\bu} M_{\hi}{}^{\pj}\d_{1\lambda}\d_{2\b}+\d_{\bu}{}^{\bv}\bar M_{\pj}{}^{\hi}\d_{2\lambda}\d_{1\b}).
\ee
This is the third example that if the bosonic subalgebras of two superalgebras $G$ and $G^\prime$ have a common part, then their fermionic generators have nontrivial anticommutators.

We now combine the two independent sets \emph{fermionic} oscillators as one set:
\e\label{twoosci3}
&&A_I=\begin{pmatrix} a_{\hi}\\c_{\pri}\end{pmatrix},\quad A^{I\dag}=\begin{pmatrix} a^{\hi\dag}\\c^{\pri\dag}\end{pmatrix},\quad
\{A_{I}, A^{J\dag}\}=\d_{I}{}^{J},\quad \{A_{I}, A_{J}\}=\{A^{I\dag}, A^{J\dag}\}=0,
\nonumber\\&&\d_{I}{}^{J}=\begin{pmatrix}\d_{\hi}{}^{\hj} & 0\\0 &\d_{\pri}{}^{\pj} \end{pmatrix},
\ee
where $I=1,\ldots, (N_1+N_3)$ is a collective index. We now are able to define
\e\label{cmbfmg}
Q_{\bu}{}^I=\begin{pmatrix} Q_{\bu}{}^{\hi}\\Q_{\bu}{}^{\pri}\end{pmatrix}=\sqrt{-k}b_{\bu}A^{I\dag},\quad
\bar Q_{I}{}^{\bu}=\begin{pmatrix} \bar Q_{\hi}{}^{\bu}\\\bar Q_{\pri}{}^{\bu}\end{pmatrix}=\sqrt{-k}A_{I}b^{\bu\dag}.
\ee
As a result, we can put the known anticommutator and commutators in the compact forms
\begin{equation}\label{anticm3}
\{Q_{\bu}{}^I,\bar Q_{J}{}^{\bv}\}=k(\d_{I}{}^J M_{\hi}{}^{\hj}+\d_{\hi}{}^{\hj}M_{I}{}^{J}),\quad
[M_{\bu}{}^{\bv},Q_{\bw}{}^I]=\d_{\bw}{}^{\bv}Q_{\bu}{}^I,\quad [M_{\bu}{}^{\bv},\bar Q_{I}{}^{\bw}]=-\d_{\bu}{}^{\bw}\bar Q_{I}{}^{\bv},
\end{equation}
where we have defined
\e\label{ospcmtt3}
M_{I}{}^{J}=\begin{pmatrix}M_{\hi}{}^{\hj} & M_{\hi}{}^{\pj}\\
\bar M_{\pj}{}^{\hi} & M_{\pri}{}^{\pj}
\end{pmatrix}.
\ee
By either requiring that the $Q_{I\hi}Q_{J\hj}Q_{K\hk}$ Jacobi identity is obeyed or using oscillator algebra, we can obtain
\e\label{cmm3}
[M_{I}{}^{J},Q_{\bu}{}^{K}]=-\d_{I}{}^K Q_{\bu}{}^{J},\quad[M_{I}{}^{J},\bar Q_{K}{}^{\bu}]=\d_{K}{}^J\bar Q_{I}{}^{\bu}.
\ee
Similarly, by either requiring that the $\bar Q_{I}{}^{\bu}Q_{\bv}{}^{J}M_{K}{}^{L}$ Jacobi identity is obeyed or using oscillator algebra, one can derive the following commutator
\e\label{sptt3}
[M_{I}{}^{J},M_{K}{}^{L}]=\d_{J}{}^K M_{I}{}^{L}-\d_{I}{}^L M_{K}{}^{J}.
\ee
We recognize that it is the Lie algebra of $U(N_1+N_3)$. Together with the algebra of $U(N_2)$, we find that (\ref{anticm3}), (\ref{cmm3}), and (\ref{sptt3}) furnish the commutation relations of $U(N|N_1+N_3)$. Notice that the ``fused" superalgebra $U(N|N_1+N_3)$ can be used to construct the $\CN=6$ theory \cite{Hosomichi:2008jb}.

\section{Generalizations of Fusing Superalgebras}\label{secGene}
In this section, we shall work out the general structure of the  superalgebra ``fused" by two superalgebras $G$ and $G^\prime$ whose bosonic parts share at least one simple factor or $U(1)$ factor, and work out some generalizations such as ``fusing" three or more superalgebras. The general commutation relations of $G$ and $G^\prime$ are given by  (\ref{slie2}) and (\ref{slie3}), respectively.

\subsection{General Structure of Superalgebras  Fused  by 2 Superalgebras}\label{secgw}

Given two superalgebras $G$ and $G^\prime$, what interests us most is that the bosonic parts of the superalgebras $G$ and $G^\prime$ share at least one simple factor or $U(1)$ factor, while we do \emph{not} identify $G$ and $G^\prime$. Schematically, we have $M^u=(M^\a,M^g)$ and $M^\pu=(M^{\a^\prime},M^g)$, with  $M^g$ the set of generators of the common bosonic part of $G$ and $G^\prime$, i.e.  $M^u\cap M^\pu=M^g\neq \emptyset$, but we \emph{exclude} the possibility that $Q_a=Q_\pa$ and $M^{\a^\prime}=M^\a=\emptyset$. Equivalently, we must require that \cite{ChenWu4}
\e
\kappa(Q_a, Q_\pb)=\omega_{a\pb}=0.
\ee
(The forms $\kappa$ are defined in Appendix \ref{srt0}; for instance, $\omega_{ab}=\kappa(Q_a,Q_b)$.)

Recall that in Section \ref{Introduction} we have defined
\begin{equation}\label{QQB}
\{Q_{a}, Q_{\pb}\}= t^{\tilde{u}}_{a\pb}M_{\tilde{u}}.
\end{equation}
(See Eq. (\ref{ntrvanticm0}).)  Comparing the first equation of (\ref{newcom}) with the above equation, we have
\begin{equation}\label{Mmix}
M_{\tilde{u}}=M_{\hl\pl}, \quad (t^{\tilde{u}})_{\bi\hi,\bj\pj}=(t^{\hl\pl})_{\bi\hi,\bj\pj}
=k\d_{\bi\bj}\delta^{\hl}_{\hi}\delta^\pl_\pj.
\end{equation}
Eqs. (\ref{nicecmtt}) and (\ref{qaqpb3}) are the other two explicit examples of (\ref{QQB}). Using (\ref{QQB}) and the third equations of (\ref{4dgdcm}) and (\ref{tensorprd1}), the $Q_aQ_bQ_\pc$ Jacobi identity (\ref{jacobi}) can be converted into
\e\label{QQQBJ}
\t^g_{ab}k_{gh} \t^{g\pd}{}_\pc Q_{\pd}+t^{\tilde{u}}_{a\pc}[M_{\tilde{u}},Q_b]+t^{\tilde{u}}_{b\pc}[M_{\tilde{u}},Q_a]=0.
\ee
Notice that the first term is a linear combination of the set of generators $Q_\pd$. We are therefore led to define
\begin{equation}\label{MMBQ}
[M_{\tilde{u}},Q_a]=t_{\tilde{u}a}{}^\pd Q_\pd.
\end{equation}
Roughly speaking, the new bosonic generators $M_{\tilde{u}}$ must ``rotate" the set of fermionic generators $Q_a$ into $Q_\pd$. This means that the last term of the right hand side of $[M_{\tilde{u}},Q_a]=t_{\tilde{u}a}{}^\pd Q_\pd+t_{\tilde{u}a}{}^d Q_d$ vanishes, i.e. $t_{\tilde{u}a}{}^d=0$. A proof on that $t_{\tilde{u}a}{}^d=0$ can be found in Ref. \cite{ChenWu4}. The last equation of (\ref{newcom}) is an example of (\ref{MMBQ}).
With Eq. (\ref{MMBQ}), the $Q_aQ_bQ_\pc$ Jacobi identity
(\ref{QQQBJ}) becomes
\begin{equation}\label{QQQBJ2}
[k_{gh}\t^g_{ab}\t^{h\pd}{}_{\pc}+t^{\tilde{u}}_{a\pc}t_{\tilde{u}b}{}^\pd
+t^{\tilde{u}}_{b\pc}t_{\tilde{u}a}{}^\pd]Q_\pd=0.
\end{equation}
This is a non-linear constraint on these structure constants. Now the key point is that we have to define $t_{\tilde{u}a}{}^\pd$ carefully so that the identity (\ref{QQQBJ2}) is obeyed. Similarly, requiring that the $Q_\pc Q_\pd Q_a$ Jacobi identity is obeyed leads us to define
\begin{equation}\label{MMBQB}
[M_{\tilde{u}},Q_\pc]=t_{\tilde{u}}{}^b{}_\pc Q_b.
\end{equation}
Namely, the new bosonic generators $M_{\tilde{u}}$ ``rotate" the set of fermionic generators $Q_\pc$ into $Q_b$. The fist equation of the second line of (\ref{newcom}) is an example of (\ref{MMBQB}). With (\ref{MMBQB}), the $Q_\pa Q_\pb Q_c$ Jacobi identity can be converted into
\e
[k_{gh}\t^g_{\pc\pd}\t^{hb}{}_{a}+t^{\tilde{u}}_{a\pc}t_{\tilde{u}}{}^b{}_\pd
+t^{\tilde{u}}_{a\pd}t_{\tilde{u}}{}^b{}_\pc]Q_b=0.
\ee
The equation in the bracket is essentially equivalent to that of (\ref{QQQBJ2}). This is also consistent with the requirement that using either $[\{Q_a,Q_b\},Q_{\pc}]$ or $[\{Q_\pc,Q_\pd\},Q_{a}]$ to calculate the structure constants $f_{ab\pc\pd}$ gives the same result
\e
f_{ab\pc\pd}=k_{gh}\t^g_{ab}\t^{h}_{\pc\pd}.
\ee

All other unknown commutators, $[M^{\tilde{u}}, M^{\tilde v}]$, $[M^u, M^{\tilde u}]$, and $[M^{\pu}, M^{\tilde u}]$ can be determined by requiring that the $M^{\tilde u}Q_aQ_\pb$, $M^uQ_aQ_\pb$, and $M^{\pu}Q_aQ_\pb$ Jacobi identities are obeyed, respectively. For instance, let us consider the $M^{\tilde u}Q_aQ_\pb$ Jacobi identity
\begin{equation}
[M^{\tilde u},\{Q_a, Q_\pb\}]-\{[M^{\tilde u},Q_a],Q_\pb\}-\{[M^{\tilde u},Q_\pb],Q_a\}=0.
\end{equation}
After some algebraic steps, we obtain the equation
\begin{equation}\label{MMmix}
(t_{\tilde v})_{a\pb}[M^{\tilde u}, M^{\tilde v}]=t^{\tilde u}{}^c{}_\pb\t^u_{ac}k_{uv}M^v+(t^{\tilde u})_{a}{}^\pc\t^\pu_{\pc\pb}k_{\pu\pv}M^\pv,
\end{equation}
which determines the commutator $[M^{\tilde u}, M^{\tilde v}]$, since $(t_{\tilde v})_{a\pb}$ is generally consisted of by invertible invariant tensors (the last equation of (\ref{Mmix}) is such an example). Eq. (\ref{pmpm}) is an explicit example which can be derived from Eq. (\ref{MMmix}). Though it seems that both $M^u$ and $M^\pu$ appear in the right hand side of (\ref{MMmix}), their common generators $M^g$ are generally absent in the right hand side of the equation \begin{equation}[M^{\tilde u}, M^{\tilde v}]=f^{\tilde u\tilde v}{}_{\tilde w}M^{\tilde w}.\end{equation} For instance, in Eq. (\ref{pmpm}), the common generators $M_{\bi\bj}$ do not appear in the right hand side. Similarly, by using Jacobi identities, we can obtain
\e\label{MMmix2}
&&(t_{\tilde v})_{a\pb}[M^u, M^{\tilde v}]=(\t^{ u}{}^\pc{}_\pb t^{\tu}_{a\pc}k_{\tu\tv}+\t^{uc}{}_{a}t^{\tu}_{c\pb}k_{\tu\tv})M^{\tv},\\
&&(t_{\tilde v})_{a\pb}[M^{\pu}, M^{\tilde v}]=(\t^{\pu c}{}_{a}t^{\tu}_{c\pb}k_{\tu\tv}+\t^{\pu}{}^\pc{}_\pb t^{\tu}_{a\pc}k_{\tu\tv})M^{\tv}.\label{MMmix3}
\ee
Eqs. (\ref{unpmnew}) and (\ref{pmnew}) are two explicit examples which can be derived from (\ref{MMmix2}) and (\ref{MMmix3}), respectively.

In summary, we have defined the commutation relations (\ref{QQB}), (\ref{MMBQ}), (\ref{MMBQB}), (\ref{MMmix}), (\ref{MMmix2}), and (\ref{MMmix3}) for fusing $G$ and $G^\prime$, by requiring that the corresponding Jacobi identities are obeyed. If all Jacobi identities of the `total" superalgebra consisting of $M_{\tilde{u}}$ \emph{and} all generators of $G$ \emph{and} $G^\prime$ are satisfied, we say that the superalgebra ``fused" by $G$ \emph{and} $G^\prime$ is closed. Notice that in fusing $G$ and $G^\prime$, we have \emph{not} introduced any fermionic generators; the set of bosonic generators $M_{\tilde{u}}$ are the only ones introduced for the fusion. An alternate approach is that one can use oscillators to construct the generators of $G$ and $G^\prime$ first, then use straightforward oscillator algebra to derive the commutation relations (\ref{QQB}), (\ref{MMBQ}), (\ref{MMBQB}), (\ref{MMmix}), (\ref{MMmix2}), and (\ref{MMmix3}), as we did in Sections \ref{secfsosp1}, \ref{secfsosp2}, and \ref{secfumun}.

We can combine the two sets of fermionic generators $Q_a$ and $Q_\pb$ into one set
\e
Q_I=\begin{pmatrix}Q_a\\Q_\pa \end{pmatrix}.
\ee
Eqs. (\ref{cmb2osci}), (\ref{2fmges}), and (\ref{cmbfmg}) are explicit examples of the above combination. As a result, all commutation relations of the ``fused" superalgebra can be put into the compact form
\begin{equation}\label{finalfusion}
\{Q_I,Q_J\}=\tau^U_{IJ}k_{UV}M^V,\quad[M^U,Q_I]=-\tau^U_{IJ}\omega^{JK}Q_K ,\quad [M^U,M^V]=f^{UV}{}_WM^W.
\end{equation}
For instance, Eqs. (\ref{anticm}), (\ref{cmm}), and (\ref{sptt}) are concrete examples of the first, second, and third equations of (\ref{finalfusion}), respectively.

The $M^gM^hQ_\pa$ and $M^g M^h Q_a$ Jacobi identities are nontrivial and always obeyed due to the fact that $M^u\cap M^\pu=M^g\neq \emptyset$ (see Appendix \ref{srt0}), so that the second and third FIs of (\ref{FI4}) are always satisfied, even one cannot fuse $G$ and $G^\prime$ into a single closed superalgebra.

\subsection{Fusing More Than Two Superalgebras}\label{secmore2}
It is straightforward to generalize the ``fusion" procedure to fuse three or more superalgebras. We may have to add new fermionic generators to fuse three or more superalgebras. Consider for example the following superalgebras
\e\label{addfm}
(G_1,G_2,G_3)=\big(U(N_1|N_2), U(N_2|N_3), U(N_3|N_4)\big).
\ee
Here $G_1$ and $G_2$, whose bosonic parts share a common part $U(N_2)$, satisfy the conditions as $G$ and $G^\prime$ do (see Sec. \ref{secGene}); $G_2$ and $G_3$ also satisfy the conditions as $G$ and $G^\prime$ of Sec. \ref{secGene} do, but their bosonic parts share a common part $U(N_3)$. However, \emph{the superalgebra $G_1$ is independent of $G_3$}, in the sense that every generator of $G_1$ commutes or anticommutes with every generator of $G_3$. Using (\ref{addfm}) to construct the 3-algebra in the  $\CN=4$ quiver gauge theory gives the quiver diagram for the gauge group \cite{ChenWu4}
\e\label{3unis}
U(N_1)-U(N_2)-U(N_3)-U(N_4).
\ee
The \emph{three} copies of multiplets are in the bifundamental representations of $U(N_1)\times U(N_2)$, $U(N_2)\times U(N_3)$, and  $U(N_3)\times U(N_4)$, respectively.

To fuse $G_1\sim G_3$, let us try to utilize their oscillator realizations. Recall that $G_1$ and  $G_2$ are constructed in terms three independent oscillators in Eqs. (\ref{oscillators20}), and their generators are given by (\ref{osciunis1}) and (\ref{osciunis2}), respectively. To construct $G_3$ in terms of oscillators, we need to introduce a \emph{fourth} independent set of oscillators
\e\label{fourth}
[d_{\check u}, d^{\check v\dag}]=\d_{\check u}{}^{\check v},\quad [d_{\check u}, d_{\check v}]=[d^{\check u\dag}, d^{\check v\dag}]=0.
\ee
Now the generators of $G_3$ can be constructed in terms of (\ref{fourth}) and the third set of oscillators of (\ref{oscillators20}):
\e\label{osciunis3}
Q_{\check u}{}^{\pri}=\sqrt{-k}d_{\check u}c^{\pri\dag},\quad \bar Q_{\pri}{}^{\check u}=\sqrt{-k}c_{\pri}d^{\check u\dag},\quad M_{\pri}{}^{\pj}=a_{\pri}a^{\pj\dag},\quad M_{\check u}{}^{\check v}=-d_{\check u}d^{\check v\dag}.
\ee
Let us now pick up three fermionic generators from $G_1$, $G_2$, and $G_3$ respectively, and consider the Jacobi identity of these 3 generators:
\e
[\{\bar Q_{\hi}{}^{\bu}, Q_{\bv}{}^{\pri}\},\bar Q_{\pk}{}^{\check u}]+[\{\bar Q_{\pk}{}^{\check u}, Q_{\bv}{}^{\pri}\},\bar Q_{\hi}{}^{\bu}]+[\{\bar Q_{\hi}{}^{\bu}, \bar Q_{\pk}{}^{\check u}\},Q_{\bv}{}^{\pri}]=0.
\ee
Without any calculating, we notice immediately that the last term must vanish since $G_1$ is independent of $G_3$. A short calculation shows that the first two terms add up to be zero. The explicit expression of the first term is given by
\e
[\{\bar Q_{\hi}{}^{\bu}, Q_{\bv}{}^{\pri}\},\bar Q_{\pk}{}^{\check u}]=\d_{\bv}{}^{\bu}\d_{\pk}{}^{\pri}(\sqrt{-k}b_{\hi}d^{\check u\dag})\equiv\d_{\bv}{}^{\bu}\d_{\pk}{}^{\pri}\bar Q_{\hi}{}^{\check u},
\ee
where we have defined a set of fermionic generators $\bar Q_{\hi}{}^{\check u}$. Similarly, the $Q_{\bu}{}^{\hi}\bar Q_{\pri}{}^{\bv}Q_{\check u}{}^{\pk}$ Jacobi identity is also obeyed, and we must introduce another set of \emph{fermionic generators} $Q_{\check u}{}^{\hi}$ defined by the equation
\e
[\{Q_{\bu}{}^{\hi}, \bar Q_{\pri}{}^{\bv}\},Q_{\check u}{}^{\pk}]=-\d_{\bu}{}^{\bv}\d_{\pri}{}^{\pk}(\sqrt{-k}d_{\check u}b^{\hi\dag})\equiv-\d_{\bu}{}^{\bv}\d_{\pri}{}^{\pk} Q_{\check u}{}^{\hi}.
\ee
Namely, we must introduce the \emph{fourth} set of \emph{fermionic generators}
\e
Q_{\check{a}}=\begin{pmatrix} \bar Q_{\hi}{}^{\check u}\\ -Q_{\check u}{}^{\hi}\end{pmatrix}
\ee
into the system. According to Sec. \ref{Introduction}, this is a typical \emph{fermionic fusion}\footnote{The definition of fermionic fusion is essentially different from that of the fusion in Sec. \ref{Introduction} (see also Sec. \ref{secGene}), since in defining the fusion, we have not introduced any new fermionic generators. The fermionic fusion may be not interesting as the fusion itself, since in principal, by adding sufficient fermionic generators, one may fuse any two or more superalgebras into a single close superalgebra. However, it is still interesting to ask that what are the \emph{minimum numbers} of fermionic generators needed for fusing two or more superalgebras into a single closed one.}, since we have introduced a set of new fermionic generators $Q_{\check{a}}$ to fuse the three superalgebras. It is easy to verify that $\bar Q_{\hi}{}^{\check u}$ and $Q_{\check u}{}^{\hi}$ obey the commutation relations of $U(N_1|N_4)$; for instance,
\e
\{Q_{\check u}{}^{\hi}, \bar Q_{\hj}{}^{\check v}\}=k(\d_{\check u}{}^{\check v}M_{\hj}{}^{\hi}+\d_{\hj}{}^{\hi}M_{\check u}{}^{\check v}).
\ee
Therefore, to obtain a fermionic fusion of $G_1\sim G_3$, it is necessarily to introduce the fermionic generators of $U(N_1|N_4)$ hence the the superalgebra $U(N_1|N_4)$ into the system.
Let us denote $U(N_1|N_4)$ as $G_4$. Then the fermionic fusion of the superalgebras $G_1\sim G_3$ (see (\ref{addfm})), is the same as the fusion of the four superalgebras
\begin{equation}\label{G1G4u}
(G_1,G_2,G_3,G_4)=\big(U(N_1|N_2), U(N_2|N_3), U(N_3|N_4),U(N_1|N_4)\big).
\end{equation}
These superalgebras form a closed ``loop"
\e\label{loopg}
&&\quad\quad G_1\nonumber\\
&&G_4\quad\quad\quad G_2\nonumber\\
&&\quad\quad G_3
\ee
in which the bosonic parts of every adjacent pair share one common part. The superalgebra fused by $G_1\sim G_4$ is $U(N_1+N_3|N_2+N_4)$. If we use (\ref{G1G4u}) to construct the  $\CN=4$ theory \cite{ChenWu4}, in accordance with (\ref{loopg}), the resulting quiver diagram for the gauge group is
\e\label{4unisu}
&&U(N_1)-U(N_2)\nonumber\\
&&\quad  |\quad\quad\quad\quad |   \nonumber\\
&&U(N_4)-U(N_3)
\ee
The \emph{four} copies of multiplets are in the bifundamental representations of $U(N_1)\times U(N_2)$, $U(N_2)\times U(N_3)$ , $U(N_3)\times U(N_4)$, and $U(N_4)\times U(N_1)$, respectively.

We see that the theory constructed by using (\ref{G1G4u}) is completely different from the one constructed by using (\ref{addfm}). However, they have the \emph{same} underlying structure, in the sense that both (\ref{addfm}) and (\ref{G1G4u}) can be ``fused" into the same superalgebra $U(N_1+N_3|N_2+N_4)$. We have to emphasize that the Lie algebra of the gauge group represented by either (\ref{3unis}) or (\ref{4unisu}) is just a proper subalgebra of the bosonic part of $U(N_1+N_3|N_2+N_4)$, \emph{not} its full bosonic part.

Also, if we select the bosonic parts of $U(N_1+N_3|N_2-\lambda)$ and $U(N_1+N_3|N_4+\lambda)$ ($\lambda=0,\ldots,N_2-1$), sharing the common factor $U(N_1+N_3)$, as the Lie algebra
of the gauge symmetry, we will obtain a set of different $\CN=4$ theories as $\lambda$ runs from $0$ to $N_2-1$; the corresponding quiver diagrams are given by
\e\label{2unis}
U(N_2-\lambda)-U(N_1+N_3)-U(N_4+\lambda),\quad \lambda=0,\ldots,N_2-1.
\ee
However, all the corresponding ``fused" superalgebras are the same, since
\e\nonumber
&&\bigg(U(N_1+N_3|N_2-\lambda)\quad {\rm fusing }\quad U(N_1+N_3|N_4+\lambda)\bigg)\\&=&U(N_1+N_3|N_2-\lambda+N_4+\lambda)
=U(N_1+N_3|N_2+N_4).
\ee
So the theories depending on $\lambda$ are not only different form each other, but also different
from the two theories constructed by using (\ref{G1G4u}) and (\ref{addfm}) respectively. But all these theories are associated with the same ``fused" superalgebra $U(N_1+N_3|N_2+N_4)$.

Let us now try to ``fuse" the superalgebras in (\ref{addfm}) in an alternative approach. On one hand, fusing $G_1$ and $G_2$ first gives $U(N_1+N_3|N_2)$. We now must fuse $U(N_1+N_3|N_2)$ and $G_3=U(N_3|N_4)$. On the other hand, fusing $G_2$ and $G_3$ first gives $U(N_3|N_2+N_4)$. Hence we must fuse $U(N_3|N_2+N_4)$ and $G_1=U(N_1|N_2)$. The two fusions must give the same final result, i.e.
\begin{equation}
\bigg(U(N_1+N_3|N_2)\quad {\rm fusing }\quad U(N_3|N_4)\bigg)=\bigg(U(N_3|N_2+N_4)\quad {\rm fusing }\quad U(N_1|N_2)\bigg).
\end{equation}
Let us look at the left hand side. We see that the common part, the Lie algebra of $U(N_3)$, is \emph{a proper subalgebra} of the Lie algebra of $U(N_1+N_3)$ of the bosonic part of $U(N_1+N_3|N_2)$, provided that $N_1\neq0$. As a result, we encounter an interesting complication. However, by using a little oscillator algebra, we find that it is necessarily to add $U(N_1|N_4)$ into the left hand side, and fusing these three superalgebras gives exactly $U(N_1+N_3|N_2+N_4)$, which also can be derived by fusing $U(N_1|N_4)$ and the two superalgebras in the right hand side.

Consider the general line-like (\emph{not} a closed loop) case of unitary superalgebras
\e\label{nunum}
(G_1,\ldots, G_{n})=\big(U(N_1|N_2), U(N_2|N_3),\ldots, U(N_{n-1}|N_{n}),U(N_{n}|N_{n+1})\big),
\ee
where the bosonic parts of any adjacent pair $G_i$ and $G_{i+1}$ ($i=1,\cdots,n-1; n\geq3)$
share a common part $U(N_{i+1})$; any pair of superalgebras are independent
if they not adjacent ($G_1$ and $G_{n}$ are also independent). They can be fused into
\e\label{finunum}
U\bigg(\sum_{k=1}^{1+[\frac{n}{2}]}N_{2k-1}\bigg|\sum_{k=1}^{[\frac{n+1}{2}]}N_{2k}\bigg),
\ee
where $[\frac{n}{2}]$ is the integer part of $\frac{n}{2}$. In (\ref{nunum}), if $n$ is \emph{even} and the bosonic parts of $G_1$ and $G_n$ share the Lie algebra of $U(N_1)=U(N_{n+1})$, then (\ref{nunum}) becomes a closed loop and the resulting ``fused" superalgebra is also (\ref{finunum}). However, if $n$ is \emph{odd}, it seems that one cannot fuse these superalgebras forming a closed loop into a single closed superalgebra. Also, their bosonic parts cannot be selected as the Lie algebra of gauge group of the $\CN=4$ theories \cite{HosomichiJD}. For example, consider the simplest case
\begin{equation}\label{G1G3cl}
(G_1,G_2,G_3)=(U(N_1|N_2), U(N_2|N_3), U(N_3|N_1)),
\end{equation}
which is a closed loop with odd number of superalgebras, in the sense that the bosonic parts of \emph{every} pair superalgebras share one common part. If we pick up three fermionic generators from $G_1$, $G_2$ and $G_3$, respectively, then their Jacobi identity cannot be obeyed. Also, if we select the bosonic parts of (\ref{G1G3cl}) as the Lie algebra of gauge group of the  $\CN=4$ theory, then at least two multiplets (out of the three multiplets) are in the dotted or undotted representation of the $SU(2)\times SU(2)$ R-symmetry group simultaneously. As a result, at least one $QQQ$ Jacobi identity cannot be obeyed, so the bosonic parts of  $G_1$, $G_2$ and $G_3$ cannot be selected as the Lie algebra of gauge group of the  $\CN=4$ theory \cite{HosomichiJD}.

Similarly, one can also consider the line-like case of orthosymplectic superalgebras,
\e\label{norsym}
&&\big(OSp(M_1|2N_1), OSp(M_1|2N_2), OSp(M_2|2N_2),\ldots,\nonumber
\\ &&\quad\quad  OSp(M_n|2N_n), OSp(M_n|2N_{n+1}), OSp(M_{n+1}|2N_{n+1})\big)
\ee
where the bosonic parts of any adjacent pair of superalgebras
share one and only one common simple factor; any pair of superalgebras are independent
if they not adjacent ($(OSp(M_1|2N_1)$ and $OSp(M_{n+1}|2N_{n+1})$ are independent as well). One can fuse them into
\e\label{fiorsym}
OSp\bigg(\sum_{k=1}^{n+1}M_{k}\bigg|2\sum_{k=1}^{n+1}N_{k}\bigg).
\ee
If the number of all superalgebras in (\ref{norsym}) is \emph{even} and the bosonic parts of the first and last superalgebras ($(OSp(M_1|2N_1)$ and $OSp(M_{n+1}|2N_{n+1})$) share one simple common factor $SO(M_1)=SO(M_{n+1})$, then (\ref{norsym}) becomes a closed loop and the resulting ``fused" superalgebra is also (\ref{fiorsym}).

If the bosonic parts of three or more superalgebras share one common part, they can be also fused into a single closed superalgebra. For example, let us consider
\e\label{mutisha1}
(G_1,\ldots,G_n)=\big(OSp(M|2N_1),\ldots, OSp(M|2N_i),\ldots, OSp(M|2N_n)\big),
\ee
i.e. the bosonic parts of the $n\geq3$ orthosymplectic superalgebras share the Lie algebra of $SO(M)$. They can be fused into
\e\label{fmutisha1}
OSp\bigg(M\bigg|2\sum_{i=1}^{n}N_i\bigg).
\ee
Similarly, one can fuse
\e\label{mutisha2}
&&(G_1,\ldots,G_n)=\big(OSp(M_1|2N),\ldots, OSp(M_i|2N),\ldots, OSp(M_n|2N)\big)\\
&&{\rm and}\quad(G_1,\ldots,G_n)=\big(U(N|N_1),\ldots, U(N|N_i),\ldots, U(N|N_n)\big)\label{mutisha3}
\ee
into
\e\label{fmutisha2}
OSp\bigg(\sum_{i=1}^n M_i\bigg|2N\bigg) \quad{\rm and}\quad U\bigg(N\bigg|\sum_{i=1}^nN_i\bigg),
\ee
respectively. However, by the same reason that the bosonic parts of (\ref{G1G3cl}) cannot be selected as the Lie algebra of gauge group of the  $\CN=4$ quiver gauge theory, the bosonic parts of (\ref{mutisha1}), (\ref{mutisha2}) or (\ref{mutisha3}) cannot be used to construct the  $\CN=4$ quiver gauge theory, though their ``fused" superalgebras (\ref{fmutisha1}) and (\ref{fmutisha2}) can be used to construct the $\CN=4$ GW theories and $\CN=5$ theories.

It is also possible to ``fuse" the superalgebras whose bosonic parts forming a more complicated mesh-like diagram. For example, consider the following seven superalgebras
\begin{eqnarray}\label{meshy}
(G_1,G_2,G_3,G_4,G_5,G_6,G_7)&=&
(OSp(M|2N_1),OSp(M|2N_2),OSp(M|2N_3),OSp(M|2N),\nonumber\\&&\quad
OSp(M_1|2N),OSp(M_2|2N),OSp(M_3|2N)).
\end{eqnarray}
The bosonic parts of $G_1\sim G_4$ share the Lie algebra of $SO(M)$, while the bosonic parts of $G_5\sim G_7$ share the Lie algebra of $Sp(2N)$. Therefore the bosonic parts of the seven superalgebras form the mesh-like diagram:
\begin{eqnarray}\label{meshy2}
& Sp(2N_1)\quad SO(M_3)\nonumber\\
&\mid\quad\quad\quad\quad\quad\mid\nonumber\\
&Sp(2N_2)-SO(M)-Sp(2N)-SO(M_2)\quad.\\
&\mid\quad\quad\quad\quad\quad\mid\nonumber\\
&Sp(2N_3)\quad SO(M_1)\nonumber
\end{eqnarray}
These seven superalgebras (\ref{meshy}) can be fused into the superalgebra
\e
OSp\bigg(M+\sum_{i=1}^3M_i\bigg|2\big(N+\sum_{i=1}^3N_i\big)\bigg),
\ee
which can be used to construct the $\CN=4$ GW theories and $\CN=5$ theories.

\section{Conclusions and Discussion}\label{conclusions}

We have developed a \emph{fusion} procedure to ``fuse" two
superalgebras $G$ and $G^\prime$, whose bosonic parts share at least one
simple factor or $U(1)$ factor, into a single closed superalgebra. The
fermionic generators of the ``fused" superalgebra are a disjoint
union of the fermionic generators of $G$ and $G^\prime$; in fusing
$G$ and $G^\prime$, one needs only to introduce a set of new
bosonic generators for closing the ``fused" superalgebra. The
generic structure of the superalgebra ``fused" by two
superalgebras has been worked out, and the fusion procedure has
been generalized so that one can fuse more than two superalgebras.
Two different methods were introduced to do the fusion. We have
constructed several classes of the ``fused" superalgebras in Sec.
\ref{secfosps}, \ref{secfumun} and \ref{secGene}. For instance, in
Sec. \ref{secfumun}, we have fused $U(N_1|N_2)$ and $U(N_2|N_3)$
into $U(N_1+N_3|N_2)$. Here the common part of the bosonic parts
of $U(N_1|N_2)$ and $U(N_2|N_3)$ is $U(N_2)$. It seems all
classical superalgebras admit ``fusions", i.e. two orthosymplectic
(unitary) superalgebras can be fused into a single closed
orthosymplectic (unitary) superalgebra, provided that certain
conditions are satisfied.

We have also generalized the \emph{fusion} procedure to a
\emph{fermionic fusion} procedure, by allowing one to add
\emph{minimum numbers} of fermionic generators as well as bosonic
generators into the system, such that two or more superalgebras
may be fused into a closed one (see Sec. \ref{secmore2}).

It is particularly interesting to note that even if two or more $\CN=4$
theories have completely different gauge groups and different
numbers of multiplets, they may have the same underlying ``fused"
superalgebra structure, in the sense that the corresponding two or more sets of the superalgebras, used to construct the 3-algebras that
generate the gauge groups, can be ``fused" into the same single
closed superalgebra, respectively.  For instance, the $\CN=4$ quiver gauge
theories whose quiver diagrams are given by (\ref{3unis}),
(\ref{4unisu}), and (\ref{2unis}), respectively,
are associated with the same ``fused" superalgebra $U(N_1+N_3|N_2+N_4)$. It would be nice to explore the physical
significance of this relationship.

We have also discovered that some superalgebras \emph{cannot} be
fused into a single closed superalgebra even if the bosonic parts of
any pair of them share one common factor: (\ref{G1G3cl}) is such
an example. Interestingly, the bosonic parts of (\ref{G1G3cl})
cannot be selected as the Lie algebra of Lie group of the $\CN=4$
theory.

We are not sure that whether all the superalgebras in our recent
work \cite{ChenWu4} used to construct the symplectic superalgebras
in the $\CN=4$ theories can be fused into single superalgebras or
not. For instance, can we fuse $(OSp(N_1|2),G_3)$ into a single
closed superalgebra? Here the common part of the bosonic parts of
the two superalgebras is $Sp(2)\cong SU(2)$. It would be nice to
achieve a complete classification of the ``fused" superalgebras.
It can be seen that the Lie algebras of the gauge groups of the
$\CN=4$ quiver gauge theories have extremely rich structures,
hence may inspire some further non-trivial physical and
mathematical problems to study.

\section{Acknowledgement}
FMC is supported by the China Postdoctoral Science Foundation through Grant No. 2012M510244.
YSW is supported in part by the US NSF through Grant No.
PHY-1068558.

\appendix
\section{A Review of the $\CN=4$ Theory Based on 3-Algebras }\label{secN4}
In this Appendix, we review the general $\CN=4$ quiver theory
constructed in terms of the double-symplectic 3-algebra or the
$\CN=4$ three-algebra \cite{ChenWu3, ChenWu4}. The generators of
the double-symplectic 3-algebra are the disjoint union of that two
sub symplectic 3-algebras, whose generators are denoted as $T_a$
and $T_\pa$, respectively, where $a=1,\cdots,2R$ and
$\pa=1,\cdots, 2S$. The symplectic 3-algebra is a complex vector
space equipped with the 3-bracket
\begin{eqnarray}\label{Symp3Bracket}
[T_I,T_J;T_K]&=&f_{IJK }{}^dT_{d}+ f_{IJK }{}^\pd T_{\pd}\\
&\equiv& g_{IJK}{}^LT_L,\nonumber
\end{eqnarray}
where $T_I$ can be a primed \emph{or} an unprimed generator, i.e.
\e\label{compact}
T_I=(T_a\!&{\rm or}\!& T_\pa).
\ee
The 3-bracket is required to satisfy the fundamental identity (FI):
\begin{equation}\label{FI}
[T_I,T_J; [T_M,T_N;T_K]]=[[T_I,T_J;T_M],T_N;
T_K]+[T_M,[T_I,T_J;T_N]; T_K]+[T_M,T_N; [T_I,T_J;T_K]].
\end{equation}
Substituting (\ref{Symp3Bracket}) into
(\ref{FI}), we see that the structure constants must obey the identity
\begin{equation}\label{FFI}
g_{MNK}{}^Og_{IJO}{}^{L}=g_{IJM}{}^Og_{ONK}{}^{L}
+g_{IJN}{}^Og_{MOK}{}^{L}+g_{IJK}{}^Og_{MNO}{}^{L}.
\end{equation}

To define two symplectic 3-algebras, we introduce two invariant anti-symmetric tensors
\e\label{forms0}
\omega_{ab}=\omega(T_a, T_b)\quad {\rm and}\quad \omega_{\pa\pb}=\omega(T_\pa, T_\pb)
\ee
into the two sub 3-algebras, respectively, and denote their inverses as $\omega^{bc}$ and $\omega^{\pb\pc}$, satisfying
$\omega_{ab}\omega^{bc}=\d^c_a$ and $\omega_{\pa\pb}\omega^{\pb\pc}=\d^\pc_\pa$. 
We will use the antisymmetric tensors $\omega$ to lower or raise the indices. The unprimed and primed vectors are required to be symplectic orthogonal, that is,
\e\label{ortho}
\omega(T_a, T_\pb)=\omega(T_\pb, T_a)=0.
\ee

Finally, we assume that the 3-brackets satisfy the two conditions
\e\label{symin2}
&&[T_I,T_J;T_K]=[T_J,T_I;T_K],\\
&&\omega({[T_I,T_J;T_K], T_L})=\omega({[T_K,T_L;T_I], T_J}).\label{bigsym}
\ee

The 3-algebra defined by Eqs. (\ref{Symp3Bracket})$-$(\ref{bigsym}) is called a \emph{double-symplectic 3-algebra} \cite{ChenWu4}.

Taking account of (\ref{symin2}) and $T_I=(T_a$ \emph{or}
$T_\pa)$, we notice that (\ref{Symp3Bracket}) gives \emph{six} independent 3-brackets. Since $T_a$ and $T_\pa$ span two symplectic sub 3-algebras respectively, we must have
\begin{equation}\label{own3brcks}
[T_a,T_b;T_c]=f_{abc}{}^d T_d\quad{\rm and} \quad
[T_\pa,T_\pb;T_\pc]=f_{\pa\pb\pc}{}^\pd T_\pd.
\end{equation}
Comparing (\ref{own3brcks}) with (\ref{Symp3Bracket}), we note that
\e\label{vanish}
f_{abc}{}^\pd=f_{\pa\pb\pc}{}^d=0.
\ee
Using (\ref{ortho}), (\ref{symin2}), (\ref{bigsym}), and (\ref{vanish}), it is not difficult to prove that
\e\label{unused0}
f_{ab\pc}{}^d=f_{\pa\pb c}{}^\pd=f_{a\pb c}{}^d=f_{b\pa\pc}{}^\pd=0.
\ee
Combining (\ref{Symp3Bracket}) and (\ref{unused0}), we learn that the rest four 3-brackets are given by
\e\label{mbrcks}
&&[T_a,T_b;T_\pc]=f_{ab\pc}{}^\pd T_\pd,\quad
[T_\pa,T_\pb;T_c]=f_{\pa\pb c}{}^d T_d,\\
&&[T_a,T_\pb;T_c]=f_{a\pb c}{}^\pd T_\pd,\quad
[T_\pa,T_b;T_\pc]=f_{ b \pa\pc}{}^dT_d.\label{mbrcks3}
\ee
On account of the symmetry conditions (\ref{symin2}) and (\ref{bigsym}), Eq.
(\ref{FFI}) may be decomposed into eight independent FIs. The
four FIs not involving $f_{a\pb c\pd}=\omega_{\pd\pe}f_{a\pb c}{}^\pe$
are given by
\e \label{FI4}&&f_{abe}{}^gf_{gfcd}+
f_{abf}{}^gf_{egcd}-f_{efd}{}^gf_{abcg}-f_{efc}{}^gf_{abdg}=0 ,\nonumber\\
&&f_{abe}{}^gf_{gf\pc\pd}+ f_{abf}{}^gf_{eg\pc\pd}-f_{ef\pd}{}^\pg
f_{ab\pc\pg}-f_{ef\pc}{}^\pg f_{ab\pd\pg}=0 ,\\
&&f_{\pa\pb e}{}^gf_{gf\pc\pd}+ f_{\pa\pb
f}{}^gf_{eg\pc\pd}-f_{ef\pd}{}^\pg
f_{\pa\pb\pc\pg}-f_{ef\pc}{}^\pg f_{\pa\pb\pd\pg}=0,\nonumber\\
&&f_{\pa\pb \pe}{}^\pg f_{\pg\prf\pc\pd}+ f_{\pa\pb
\prf}{}^\pg f_{\pe\pg\pc\pd}-f_{\pe\prf\pd}{}^\pg
f_{\pa\pb\pc\pg}-f_{\pe\prf\pc}{}^\pg f_{\pa\pb\pd\pg}=0.
\nonumber\ee
The other four FIs involving $f_{a\pb c\pd}$ are the follows
\e \label{FI5}&&f_{a\pc b}{}^\pd f_{ef\pd\pg}=
f_{efa}{}^df_{d\pc b\pg}+f_{ef\pc}{}^\pd f_{a\pd b\pg}+f_{efb}{}^df_{a\pc d\pg} ,\nonumber\\
&&f_{a\pc b}{}^\pd f_{e\prf g\pd}=
f_{e\prf a}{}^\pd f_{\pd\pc bg}+f_{e\prf}{}^d{}_\pc f_{adbg}+f_{e\prf b}{}^\pd f_{a\pc g\pd} ,\\
&&f_{a\pc b}{}^\pd f_{\pe\prf\pd\pg}=
f_{\pe\prf a}{}^df_{d\pc b\pg}+f_{\pe\prf\pc}{}^\pd f_{a\pd b\pg}+f_{\pe\prf b}{}^df_{a\pc d\pg} ,\nonumber\\
&&f_{a\pc}{}^d{}_\pg f_{e\prf d\pb}=
f_{e\prf a}{}^\pd f_{\pd\pc\pg\pb}+f_{e\prf}{}^d{}_\pc f_{ad\pg\pb}+f_{e\prf}{}^d{}_\pg f_{a\pc d\pb}.
\nonumber\ee

We assume that the $\CN=4$ action is invariant under the transformation \cite{ChenWu3}
\e\label{glbtran1}
\delta_{\tilde\Lambda}\Phi=\Lambda^{ab}[T_a,T_b;\Phi]+\Lambda^{\pa\pb}[T_\pa,T_\pb;\Phi],
\ee
where the 3-algebra valued superfield $\Phi$ can be an untwisted superfield $\Phi=\Phi^a_AT_a$ \emph{or} a twisted superfield $\Phi=\Phi^\pa_\DA T_\pa$. The infinitesimal parameters
$\Lambda^{ab}$ and $\Lambda^{\pa\pb}$ are independent of
superspace coordinates. The symmetry (\ref{glbtran1}) will be gauged later. One may try to add the term
\e\label{hete}
\Lambda^{a\pb}[T_a,T_\pb;\Phi]\equiv \delta_{\tilde\Lambda_3}\Phi
\ee to the right hand side of (\ref{glbtran1}).
However, using the first equation of (\ref{mbrcks3}), we obtain
\e
\delta_{\tilde\Lambda_3}\Phi_A=\Lambda^{a\pb}[T_a,T_\pb;\Phi^c_AT_c]=(\Lambda^{a\pb}f_{a\pb c}{}^{\pd}\Phi^c_A) T_{\pd} .
\ee
The most right hand side indicates that $\delta_{\tilde\Lambda_3}\Phi_A\neq (\d_{\tilde\Lambda_3}\Phi)^c_AT_c$, conflicting with the assumption $\Phi_A=\Phi_A^aT_a$. We therefore must require that $\delta_{\tilde\Lambda_3}\Phi_A=0$. This can be fulfilled by setting either $\Lambda^{a\pb}=0$ \emph{or} $f_{a\pb c}{}^\pd=0$.

\begin{itemize}
\item If we set $\Lambda^{a\pb}=0$, then Eq. (\ref{hete}) does not play any role in constructing the theory;
only the symmetry defined by (\ref{glbtran1}) will be gauged.
\item If we set $f_{a\pb c}{}^\pd=0$, then (\ref{bigsym}) implies that $f_{a\pb\pd}{}^c=0$. As a result, we have $\delta_{\tilde\Lambda_3}\Phi_\DA=0$ as well. Note that after setting $f_{a\pb c}{}^{\pd}=f_{b\pa\pd}{}^{c}=0$, the four FIs (\ref{FI5}) are  satisfied automatically.
We call the new 3-algebra obtained from the
double-symplectic 3-algebra by setting $f_{a\pb c}{}^{\pd}=f_{b\pa\pd}{}^{c}=0$  an \emph{$\CN=4$ three-algebra}.

\end{itemize}

The antisymmetric tensor $\omega_{cd}$ is invariant under the transformations:
\e\label{symin20}
&&\delta_{\tilde\Lambda_1}\omega_{cd}=\Lambda^{ab}(f_{abc}{}^e\omega_{ed}
+f_{abd}{}^e\omega_{ce})=0,\\&&
\label{symin21}
\delta_{\tilde\Lambda_2}\omega_{cd}=\Lambda^{\pa\pb}(f_{\pa\pb c}{}^e\omega_{ed}+f_{\pa\pb d}{}^e\omega_{ce})=0.
\ee
Eqs (\ref{symin20}) and (\ref{symin21}) are nothing but $f_{abcd}=f_{abdc}$ and $f_{\pa\pb cd}=f_{\pa\pb dc}$, respectively. Similarly, by considering the invariance of $\omega_{\pc\pd}$, we obtain $f_{\pa\pb\pc\pd}=f_{\pa\pb\pd\pc}$ and $f_{ab\pc\pd}=f_{ab\pd\pc}$.  Note that these equations are consistent with Eq. (\ref{bigsym}). In the case of double symplectic 3-algebra, using (\ref{ortho}), it is not difficult to prove that $\omega$ are also invariant under the transformation (\ref{hete}), i.e.
\e\label{symin22}
\delta_{\tilde\Lambda_3}\omega_{cd}=\delta_{\tilde\Lambda_3}\omega_{\pc\pd}=0.
\ee
Note that in proving (\ref{symin22}), we have not set $\Lambda^{a\pb}=0$.
In the case of $\CN=4$ three-algebra, Eqs. (\ref{symin22}) are satisfied automatically due to the fact that $f_{a\pb c}{}^{\pd}=f_{b\pa\pd}{}^{c}=0$.

Plugging $\Phi=\Phi^a_AT_a$ and $\Phi=\Phi^\pa_\DA T_\pa$ into (\ref{glbtran1}), respectively, we see that only \emph{four} structure constants\footnote{Since $f_{ab\pc\pd}=f_{\pc\pd ab}$ (see
(\ref{symfs})), there are only three independent
structure constants.}
\e\label{4strc} f_{abc}{}^d,\quad f_{\pa\pb\pc}{}^\pd,\quad
f_{ab\pc}{}^\pd\quad {\rm and}\quad f_{\pa\pb c}{}^d. \ee
are needed in defining the symmetry transformation. Indeed, later we will see that only the above four structure constants appear in the
action and the law of supersymmetry transformations (see (\ref{LN4}) and
(\ref{SUSY4})), while $f_{a\pb c}{}^\pd$ and $f_{ b \pa\pc}{}^d$ (the two structure constants of the rest two 3-brackets (\ref{mbrcks3}))
do not appear in the action at all.

In summary, the four structure constants (\ref{4strc}) enjoy the following symmetry properties \cite{ChenWu3}
\begin{eqnarray}\label{symfs}
&&f_{abcd}=f_{bacd}=f_{badc}=f_{cdab},\nonumber\\
&&f_{ab\pc\pd}=f_{ba\pc\pd}=f_{ba\pd\pc}=f_{\pc\pd ab},\\
&&f_{\pa\pb\pc\pd}=f_{\pb\pa\pc\pd}=f_{\pb\pa\pd\pc}=f_{\pc\pd\pa\pb}.\nonumber
\end{eqnarray}
To guarantee the positivity of theory, they are required to obey the reality conditions \cite{ChenWu3}
\begin{equation}\label{rltcndtn}f^{*a}{}_b{}^c{}_d=f^{b}{}_a{}^d{}_c,
\quad f^{*\pa}{}_\pb{}^c{}_d=f^{\pb}{}_\pa{}^d{}_c,\quad
f^{*\pa}{}_\pb{}^\pc{}_\pd=f^{\pb}{}_\pa{}^\pd{}_\pc.
\end{equation}
To achieve the closure of the $\CN=4$ algebra, one must
impose the linear constraints \cite{ChenWu4}
\begin{equation}\label{Constr3}
f_{(abc)d}=0\quad {\rm and} \quad f_{(\pa\pb\pc)\pd}=0.
\end{equation}

It is natural to require the three independent structure constants to be invariant under the symmetry transformation (\ref{glbtran1}), i.e.
\e\label{FI6}
\d_{\tilde\Lambda}f_{abcd}=\d_{\tilde\Lambda}f_{ab\pc\pd}=\d_{\tilde\Lambda} f_{\pa\pb\pc\pd}=0,
\ee
A short calculation shows that Eqs. (\ref{FI6}) are equivalent to the
four FIs (\ref{FI4}). Therefore Eqs. (\ref{FI6}) do not involve
the rest four FIs (\ref{FI5}) at all.

Using the double-symplectic 3-algebra or the $\CN=4$ three-algebra, we have been able to construct the $\CN=4$ quiver gauge theory in a superspace approach \cite{ChenWu3}. In the theory, the un-twisted multiplets $(Z^a_A, \p^a_\DA)$ and the twisted multiplets
$(Z^\pa_\DA, \p^\pa_A)$ obey the following reality conditions
\e
\bar Z^A_a&=&\omega_{ab}\ep^{AB}Z^b_B,\quad \bar
\p^\DA_a=\omega_{ab}\ep^{\DA\DB}Z^b_\DB,\\
\bar Z^\DA_\pa&=&\omega_{\pa\pb}\ep^{\DA\DB}Z^\pb_\DB,\quad \bar
\p^A_\pa=\omega_{\pa\pb}\ep^{AB}Z^\pb_B,
\ee
where $A,\DA=1,2$ are the undotted and dotted indices of
the $SU(2)\times SU(2)$ R-symmetry group, respectively.
The $\CN=4$ Lagrangian is given
by
\e\label{LN4}\CL&=&\frac{1}{2}(-D_\mu\bar{Z}^A_aD^\mu
Z^a_A-D_\mu\bar{Z}^\DA_\pa D^\mu Z^\pa_\DA+i\bp^\DA_a\g^\mu
D_\mu\p^a_\DA+i\bp^A_\pa\g^\mu
D_\mu\p^\pa_A)\nonumber\\&&-\frac{i}{2}(f_{acbd}Z^a_AZ^{Ab}\p^c_\DB\p^{\DB
d}+ f_{\pa\pc\pb\pd}Z^\pa_\DA Z^{\DA\pb}\p^\pc_B\p^{B\pd
})\nonumber\\&&+\frac{i}{2} f_{ab\pc\pd}(Z^a_AZ^b_B\p^{A\pc}\p^{
B\pd}+Z^{\pc}_\DA Z^{\pd}_\DB\p^{\DA a}\p^{\DB b}+4Z^a_A
Z^{\DB\pd}\p^b_\DB\p^{A\pc})\nonumber\\&&
+\frac{1}{2}\epsilon^{\mu\nu\lambda}(f_{abcd}A_\mu^{ab}\partial_\nu
A_\lambda^{cd}+\frac{2}{3}f_{abc}{}^gf_{gdef}A_\mu^{ab}A_\nu^{cd}A_\lambda^{ef})\nonumber\\
&&+\frac{1}{2}\epsilon^{\mu\nu\lambda}(f_{\pa\pb\pc\pd}A_\mu^{\pa\pb}\partial_\nu
A_\lambda^{\pc\pd}+\frac{2}{3}f_{\pa\pb\pc}{}^\pg
f_{\pg\pd\pe\prf}A_\mu^{\pa\pb}A_\nu^{\pc\pd}A_\lambda^{\pe\prf})\nonumber\\
&&+\epsilon^{\mu\nu\lambda}(f_{ab\pc\pd}A_\mu^{ab}\partial_\nu
A_\lambda^{\pc\pd}+f_{abc}{}^g
f_{gd\pe\prf}A_\mu^{ab}A_\nu^{cd}A_\lambda^{\pe\prf}+f_{ab\pc}{}^\pg
f_{\pg\pd\pe\prf}A_\mu^{ab}A_\nu^{\pc\pd}A_\lambda^{\pe\prf})\nonumber\\
&&+\frac{1}{12}(f_{abcg}f^g{}_{def}Z^{Aa}Z^b_BZ^{B(c}Z^{d)}_CZ^{Ce}Z^f_A
+f_{\pa\pb\pc\pg}f^\pg{}_{\pd\pe\prf}Z^{\DA\pa}Z^\pb_\DB
Z^{\DB(\pc}Z^{\pd)}_\DC Z^{\DC\pe}Z^\prf_\DA)\nonumber\\
&&-\frac{1}{4}(f_{ab \pc\pg}f^\pg{}_{\pd ef}Z^{\DA\pc}Z^\pd_\DA
Z^b_D Z^{Df}Z^a_C Z^{Ce}+f_{\pa\pb
cg}f^g{}_{d\pe\prf}Z^{Ac}Z^d_AZ^\pb_\DD Z^{\DD\prf}Z^\pa_\DC
Z^{\DC\pe}),\nonumber\\ \ee
where the gauge fields and the covariant derivatives are defined as
\e\label{n4cov} D_\mu Z^A_d &=&
\partial_\mu Z^A_d -\tilde A_\mu{}^c{}_dZ^A_c ,\quad
\tilde A_\mu{}^c{}_d=A^{ab}_\mu f_{ab}{}^c{}_d+A^{\pa\pb}_\mu
f_{\pa\pb}{}^c{}_d ,\\ \nonumber D_\mu Z^\DA_\pd &=&
\partial_\mu Z^\DA_\pd -\tilde A_\mu{}^\pc{}_\pd Z^\DA_\pc ,\quad
\tilde A_\mu{}^\pc{}_\pd=A^{\pa\pb}_\mu
f_{\pa\pb}{}^\pc{}_\pd+A^{ab}_\mu f_{ab}{}^\pc{}_\pd. \ee
Here $A^{ab}_\mu$ and $A^{\pa\pb}_\mu$ are independent Hermitian tensors, provided that the two sub 3-algebras are not identical.
The $\CN=4$ supersymmetry transformations read
\e \label{SUSY4}&&\delta Z^a_A=i\ep_A{}^\DA\p^a_\DA,\nonumber\\
&&\delta Z^\pa_\DA=i\ep^\dag_\DA{}^A\p^\pa_A,\nonumber\\
&&\delta\p^\pa_A=-\g^\mu D_\mu
Z^\pa_\DB\ep_A{}^\DB-\frac{1}{3}f^\pa{}_{\pb\pc\pd}Z^\pb_\DB
Z^{\DB\pc}Z^\pd_\DC\ep_A{}^\DC+f^\pa{}_{\pb cd}Z^\pb_\DA
Z^{Bc}Z^d_A\ep_B{}^\DA, \nonumber\\
&&\delta\p^a_\DA=-\g^\mu D_\mu
Z^a_B\ep^\dag_\DA{}^B-\frac{1}{3}f^a{}_{bcd}Z^b_B
Z^{Bc}Z^d_C\ep^\dag_\DA{}^C+f^a{}_{b \pc\pd}Z^b_A
Z^{\DB\pc}Z^\pd_\DA\ep^\dag_\DB{}^A,\nonumber\\
&&\delta\tilde A_\mu{}^c{}_d=i\ep^{A\DB}\g_\mu\p^b_\DB
Z^a_Af_{ab}{}^c{}_d+i\ep^{\dag\DA B}\g_\mu\p^\pb_BZ^\pa_\DA
f_{\pa\pb}{}^c{}_d,\nonumber\\
&&\delta\tilde A_\mu{}^\pc{}_\pd=i\ep^{A\DB}\g_\mu\p^b_\DB
Z^a_Af_{ab}{}^\pc{}_\pd+i\ep^{\dag\DA B}\g_\mu\p^\pb_BZ^\pa_\DA
f_{\pa\pb}{}^\pc{}_\pd,\ee
where the supersymmetry parameter $\ep_{A}{}^{\dot{B}}$ obeys the following reality condition
\begin{equation}\label{n4para}
\ep^{\dag}{}_{\dot{A}}{}^{B}=
-\epsilon^{BC}\epsilon_{\dot{A}\dot{B}}\ep_{C}{}^{\dot{B}}.
\end{equation}
The closure of the above $\CN=4$
algebra has been verified in Ref. \cite{ChenWu3}.

One can generalize the construction of this Appendix
by introducing a symplectic 3-algebra containing
\emph{three or more} ($n\geq 3$) symplectic sub 3-algebras, and by letting that
$n$ multiplets take values in these $n$ sub 3-algebras respectively.
One then can realize these $n$ sub 3-algebras in terms of $n$ superalgebras respectively.

\section{A Review of the Superalgebra Realization}\label{srt0}
In this Appendix, we review the superalgebra realization of the \emph{four} sets of 3-brackets
(\ref{own3brcks}) and (\ref{mbrcks}) and the \emph{four} sets of
FIs (\ref{FI4}); we also comment on the rest two
3-brackets (\ref{mbrcks3}) and  four sets of FIs (\ref{FI5}) \cite{ChenWu4}.

As we mentioned in Section \ref{Introduction}, we used two superalgebras $G$ and $G^\prime$ to realize the two sub algebras of the double-symplectic 3-algebras \cite{ChenWu4}. Here $G$ and $G^\prime$ are given by
\begin{equation}\label{slie2} [M^u, M^v]=f^{uv}{}_wM^w,\quad [M^u,
Q_a]=-\t^u_{ab}\omega^{bc}Q_c,\quad \{Q_a,Q_b\}=\t^u_{ab}k_{uv}M^v,
\end{equation}
and
\begin{equation}\label{slie3} [M^\pu, M^\pv]=f^{\pu\pv}{}_\pw M^\pw,\quad [M^\pu,
Q_\pa]=-\t^\pu_{\pa\pb}\omega^{\pb\pc}Q_\pc,\quad
\{Q_\pa,Q_\pb\}=\t^\pu_{\pa\pb}k_{\pu\pv}M^\pv,
\end{equation}
respectively, where $a=1,\cdots,2R$ and $\pa=1,\cdots, 2S$. The
invariant antisymmetric tensors are defined as
\e\label{antitensors}
\omega_{ab}=\kappa(Q_a,Q_b),\quad
\omega_{\pa\pb}=\kappa(Q_\pa,Q_\pb),
\ee
and their inverses are
denoted as $\omega^{ab}$ and $\omega^{\pa\pb}$  satisfying
$\omega^{ab}\omega_{bc}=\d^a_c$ and
$\omega^{\pa\pb}\omega_{\pb\pc}=\d^\pa_\pc$. We will use $\omega$
to raise or lower indices. The invariant symmetric forms are
defined as $k^{uv}=-\kappa(M^u,M^v)$ and $
k^{\pu\pv}=-\kappa(M^\pu,M^\pv)$; their inverse are denoted as
$k_{uv}$ and $k_{\pu\pv}$, satisfying $k_{uv}k^{vw}=\d^w_u$ and
$k_{\pu\pv}k^{\pv\pw}=\d^\pw_\pu$. The forms $\kappa$ are
invariant \cite{GaWi,FSS} in the sense that
\begin{equation}\label{compatible0}
\kappa([A,B\},C)=\kappa(A,[B,C\}),\quad
\kappa([A^\prime,B^\prime\},C^\prime)=\kappa(A^\prime,[B^\prime,C^\prime\}),
\end{equation}
where $A=Q_a$ \emph{or} $M^u$, and $A^\prime=Q_\pa$ \emph{or} $M^\pu$.

If we  set
\e\label{eqgene}
T_a\doteq Q_a,\quad
T_\pa\doteq Q_\pa,
\ee
the four 3-brackets can be constructed in terms of the double graded commutators
\e\label{4dgdcm}
&&[T_a, T_b; T_c]\doteq [\{Q_a, Q_b\}, Q_c],\quad [T_\pa, T_\pb; T_\pc]\doteq [\{Q_\pa, Q_\pb\}, Q_\pc],\nonumber\\
&&[T_a,T_b;T_\pc]\doteq [\{Q_a, Q_b\}, Q_\pc], \quad
[T_\pa,T_\pb;T_c]\doteq [\{Q_\pa, Q_\pb\}, Q_c]. \ee The right
hand sides of the last two equations of (\ref{4dgdcm}) are
required to satisfied two crucial conditions. First, in order that
there are nontrivial interactions between the twisted and
untwisted multiplets, one must require that \e\label{2p2u}
f_{ab\pc}{}^\pd\neq0,\quad f_{\pa\pb c}{}^d\neq0. \ee Secondly, in
accordance with Eqs. (\ref{unused0}), we must require that
\e\label{vanish2} f_{ab\pc}{}^d=f_{\pa\pb c}{}^\pd =0. \ee

In Ref. \cite{ChenWu4}, we have proved that if the bosonic parts
of $G$ and $G^\prime$ share at least one simple or $U(1)$ factor,
the requirements (\ref{2p2u}) and (\ref{vanish2}) can be
fulfilled, provided that the common bosonic part of $G$ and
$G^\prime$ is not a center of $G$ and $G^\prime$. Denoting the
generators of the common bosonic part as $M^g$, i.e.
schematically, $M^g=M^u\cap M^\pu$, we have\footnote{More
generally, one can decompose $M^\pu$ into
$M^\pu=(M^{\a^\prime},\tilde M^g)$, where $\tilde M^g=T^g{}_hM^h$,
with $T^g{}_h$ a \emph{complex} non-singular linear transformation
matrix \cite{ChenWu4}. Here we set $T^g{}_h=\d^g{}_h$ for
simplicity.} \e\label{dcboso} M^u=(M^\a,M^g),\quad
M^\pu=(M^{\a^\prime},M^g). \ee (Here $\a$ is not an index of
spacetime spinor. We hope this will not cause any confusion.) And
we assume that we do \emph{not} identify the two superalgebras $G$
and $G^\prime$: schematically, we exclude the possibility that
$Q_a=Q_\pa$ and $M^{\a^\prime}=M^\a=\emptyset$. Hence it is
natural to require that
\begin{equation}
\label{noncross}[M^\a, Q_\pa]=[M^{\a^\prime}, Q_a]=0.
\end{equation}

With the decompositions (\ref{dcboso}), the anticommutators in
(\ref{slie2}) and (\ref{slie3}) can be written as \begin{equation}\label{dcanti}
\{Q_a,Q_b\}=\t^\a_{ab}k_{\a\b}M^\b +\t^g_{ab}k_{gh}M^h,\quad
\{Q_{\pa},Q_{\pb}\}=\t^\gap_{\pa\pb}k_{\gap\gbp}M^\gbp
+\t^g_{\pa\pb}k_{gh}M^h, \end{equation} where we have decomposed the
invariant quadratic form $k_{uv}$ as $k_{uv}=(k_{\a\b}, k_{gh})$.
Using (\ref{slie2}), (\ref{slie3}), and (\ref{dcanti}), the
structure constants of 3-brackets in (\ref{4dgdcm}) can be easily
read off; they are given by the tensor products
\e\label{tensorprd1} f_{abcd}=k_{uv}\t^u_{ab}\t^v_{cd}, \quad
f_{\pa\pb\pc\pd}=k_{\pu\pv}\t^\pu_{\pa\pb}\t^\pv_{\pc\pd}, \quad
f_{ab\pc\pd}= k_{gh}\t^g_{ab}\t^h_{\pc\pd}, \ee where we have used
(\ref{noncross}). By the $M^\a M^gQ_\pa$, $M^{\a^\prime} M^gQ_a$,
and $M^{\a^\prime} M^\pa Q_a$ Jocobi identities, we learn that
\e\label{trivialcm} [M^\a, M^g]=[M^{\a^\prime}, M^g]=[M^\a,
M^{\a^\prime}]=0. \ee The structure constants (\ref{tensorprd1})
posses the desired symmetry properties (\ref{symfs}) and obey the
real conditions (\ref{rltcndtn}). The $Q_aQ_bQ_c$ Jacobi identity
of (\ref{slie2}) implies that the first equation of
(\ref{Constr3}) is obeyed, i.e. $f_{(abc)d}=0$. Similarly,
$f_{(\pa\pb\pc)\pd}=0$ is equivalent to the $Q_\pa Q_\pb Q_\pc$
Jacobi identity of (\ref{slie3}).

As for the four sets of FIs in (\ref{FI4}), one can prove that they are equivalent to the
$M^uM^vQ_a$, $M^uM^vQ_\pa$, $M^g M^h Q_a$ and $M^g M^h
Q_\pa$ Jacobi identities, respectively. For
instance, using Eqs. (\ref{4dgdcm}), one of equations in (\ref{FI}) can be converted into
\begin{eqnarray}\label{n4FISP}
&&[\{Q_a,Q_b\},[\{Q_c,Q_d\},Q_\pa]]\nonumber\\&=&[\{[\{Q_a,Q_b\},Q_c],Q_d\},
Q_\pa]+[\{Q_c,[\{Q_a,Q_b\},Q_d]\},
Q_\pa]\nonumber\\&&+[\{Q_c,Q_d\},[\{Q_a,Q_b\},Q_\pa]].
\end{eqnarray}
A short calculation shows that it is equivalent to the second FI
of (\ref{FI4}). On the other hand, using (\ref{slie2}),
(\ref{dcanti}), (\ref{noncross}), and (\ref{trivialcm}), we can
convert (\ref{n4FISP})
 into the $M^gM^hQ_\pa$ Jacobi identity of (\ref{slie3})
\e\label{n4MMQjcb}
\t^g_{ab}\t^h_{cd}([M_h,[M_g,Q_\pa]]-[M_g,[M_h,Q_\pa]]+[[M_g,
M_h],Q_\pa])=0.\ee

In this realization, the Lie algebra of gauge group is the bosonic
subalgebras of the superalgebras (\ref{slie2}) \emph{and}
(\ref{slie3}); specifically, it is spanned by the set of
generators \begin{equation}\label{drcsm1}M^m=({M^\a, M^g,
M^{\a^\prime}}).\end{equation} The representations of the bosonic
subalgebras of (\ref{slie2}) and (\ref{slie3}) are determined by
the fermionic generators $Q_a$ and $Q_\pa$, respectively. The
classification of the gauge groups of the $\CN=4$ quiver gauge
theories can be found in Ref. \cite{ChenWu4, HosomichiJD,
MFM:Aug09}. In particular, in Ref. \cite{ChenWu4}, the authors
have able to construct a number of classes of $\CN=4$
theories with new gauge groups, using the approach described in this appendix.

Let us now comment on the two 3-brackets (\ref{mbrcks3}) and the
four FIs (\ref{FI5}). In the case of double-symplectic 3-algebra,
if $G$ and $G^\prime$ can be `fused' into a closed superalgebra,
one can construct the rest two 3-brackets Eqs. (\ref{mbrcks3}) in
analogue to Eqs. (\ref{4dgdcm}), i.e. \e\label{unused3b}
[T_a,T_\pb;T_c]\doteq[\{Q_a, Q_\pb\}, Q_c],\quad
[T_\pa,T_b;T_\pc]\doteq[\{Q_\pa, Q_b\}, Q_\pc]. \ee In summary, we
have \e\label{alldb} T_I\doteq Q_I,\quad[T_I,T_J;T_K]\doteq
[\{Q_I, Q_J\}, Q_K], \ee where \e\label{compactQ} Q_I=(Q_a\!&{\rm
or}\!& Q_\pa). \ee Note that (\ref{bigsym}) is obeyed by the
construction \e\label{bigsym2}
\omega([T_I,T_J;T_K],T_L)\doteq\kappa([\{Q_I,Q_J\},Q_K],Q_L), \ee
and one can also prove that Eqs. (\ref{unused0}) and
(\ref{vanish}) are obeyed \cite{ChenWu4}.

Recall that in Sec. (\ref{Introduction}), in ``fusing" $G$ and
$G^\prime$, we have defined the anticommutator of $Q_a$ and
$Q_\pb$ as (see Eq. (\ref{ntrvanticm0})) \e\label{mtm}
\{Q_a,Q_\pb\}=t^{\tu}_{a\pb}M_{\tu}, \ee where $M_{\tu}$ are a set
of bosonic generators, and $t^{\tu}_{a\pb}$ are structure
constants of the anticommutator. The structure constants of the
 commutators involving $M_{\tu}$ can be found in Sec. \ref{secGene}.
 Now one can prove that the four FIs (\ref{FI5})
  involving $f_{a\pb c\pd}$ are equivalent to the four Jacobi
  identities
  relating the bosonic generators $M_{\tu}$ defined in Eq. (\ref{mtm}).
  In summary, using (\ref{alldb}), one can construct
the FI (\ref{FI}) as follows \e [\{Q_I,Q_J\},
[\{Q_M,Q_N\},Q_K]]&=&[\{[\{Q_I,Q_J\},Q_M],Q_N\},
Q_K]+[\{Q_M,[\{Q_I,Q_J\},Q_N]\}, Q_K]\nonumber\\&&+[\{Q_M,Q_N\},
[\{Q_I,Q_J\},Q_K]],\label{FIS} \ee which can be converted into the
$MMQ$ Jacobi Identities of the fused superalgebra. By
(\ref{alldb}) and (\ref{FIS}), we learn that the double-symplectic
3-algebra indeed can be constructed in terms of the fused
superalgebra. However, if the superalgebras $G$ and $G^\prime$
cannot be fused into a closed superalgebra, we are not sure
whether one can construct (\ref{mbrcks3}) and (\ref{FI5}) in terms
of $G$ and $G^\prime$. It would be nice to answer this question.

In the case of $\CN=4$ three-algebra, the two 3-brackets vanish
identically: $[T_a, T_\pb, T_c]=[T_\pa, T_b, T_\pc]=0$. As a
result, they cannot be constructed in terms of the double graded
commutators (\ref{unused3b}) of the ``fused" superalgebra, since
the structure constants of $[\{Q_a, Q_\pb\}, Q_c]=f_{a\pb c}{}^\pd
Q_\pd$ and $[\{Q_a, Q_\pb\}, Q_\pc]=f_{a\pb \pc}{}^d Q_d$ do
\emph{not} vanish on account of the $Q_aQ_\pb Q_c$ Jacobi identity
and the $Q_aQ_\pb Q_\pc$ Jacobi identity, respectively. If both
$G$ and $G^\prime$ are unitary superalgebras \emph{or}
orthosymplectic superalgebras, by direct calculation (without
consulating the Jacobi identities), one can show that both
$[\{Q_a, Q_\pb\}, Q_c]$ and $[\{Q_a, Q_\pb\}, Q_\pc]$ are not zero
(see Sec. \ref{secfosps} and Sec. \ref{secfumun}).

\section{Conventions and Useful Identities}\label{Identities}
The conventions and useful identities are adopted from our previous
paper \cite{ChenWu3}.
\subsection{Spinor Algebra}
In $1+2$ dimensions, the gamma matrices are defined as
\begin{equation}
(\gamma_{\mu})_{\alpha}{}^\gamma(\gamma_{\nu})_{\gamma}{}^\beta+
(\gamma_{\nu})_{\alpha}{}^\gamma(\gamma_{\mu})_{\gamma}{}^\beta=
2\eta_{\mu\nu}\delta_{\alpha}{}^\beta.
\end{equation} For the metric we
use the $(-,+,+)$ convention. The gamma matrices in the Majorana
representation can be defined in terms of Pauli matrices:
$(\gamma_{\mu})_{\alpha}{}^\beta=(i\sigma_2, \sigma_1, \sigma_3)$,
satisfying the important identity
\begin{equation}
(\gamma_{\mu})_{\alpha}{}^\gamma(\gamma_{\nu})_{\gamma}{}^\beta
=\eta_{\mu\nu}\delta_{\alpha}{}^\beta+\varepsilon_{\mu\nu\lambda}(\gamma^{\lambda})_{\alpha}{}^\beta.
\end{equation}
We also define
$\varepsilon^{\mu\nu\lambda}=-\varepsilon_{\mu\nu\lambda}$. So
$\varepsilon_{\mu\nu\lambda}\varepsilon^{\rho\nu\lambda} =
-2\delta_\mu{}^\rho$. We raise and lower spinor indices with an
antisymmetric matrix
$\epsilon_{\alpha\beta}=-\epsilon^{\alpha\beta}$, with
$\epsilon_{12}=-1$. For example,
$\psi^\alpha=\epsilon^{\alpha\beta}\psi_\beta$ and
$\gamma^\mu_{\alpha\beta}=\epsilon_{\beta\gamma}(\gamma^\mu)_\alpha{}^\gamma
$, where $\psi_\beta$ is a Majorana spinor. Notice that
$\gamma^\mu_{\alpha\beta}=(\mathbbm{l}, -\sigma^3, \sigma^1)$ are
symmetric in $\alpha\beta$. A vector can be represented by a
symmetric bispinor and vice versa:
\begin{equation}
A_{\alpha\beta}=A_\mu\gamma^\mu_{\alpha\beta},\quad\quad A_\mu=-\frac{1}{2}\gamma^{\alpha\beta}_\mu A_{\alpha\beta}.
\end{equation}
We use the following spinor summation convention:
\begin{equation}
\psi\chi=\psi^\alpha\chi_\alpha,\quad\quad
\psi\gamma_\mu\chi=\psi^\alpha(\gamma_{\mu})_{\alpha}{}^\beta\chi_\beta,
\end{equation}
where $\psi$ and $\chi$ are anti-commuting Majorana spinors. In
$1+2$ dimensions the Fierz transformation reads
\begin{eqnarray}
(\lambda\chi)\psi &=& -\frac{1}{2}(\lambda\psi)\chi -\frac{1}{2}
(\lambda\gamma_\nu\psi)\gamma^\nu\chi.
\end{eqnarray}

\subsection{$SU(2)\times SU(2)$ Identities}\label{SO4}
We define the 4 sigma matrices as
\begin{equation}\label{pulim}
\sigma^a{}_A{}^{\dot{B}}=(\sigma^1,\sigma^2,\sigma^3,i\mathbbm{l}),
\end{equation}
by which one can establish a connection between the $SU(2)\times
SU(2)$ and $SO(4)$ group. These sigma matrices satisfy the following
Clifford algebra:
\begin{eqnarray}
\sigma^a{}_{A}{}^{\dot{C}}\sigma^{b\dag}{}_{\dot{C}}{}^B+
\sigma^b{}_{A}{}^{\dot{C}}\sigma^{a\dag}{}_{\dot{C}}{}^B=2\delta^{ab}\delta_A{}^B,\\
\sigma^{a\dag}{}_{\dot{A}}{}^{C}\sigma^{b}{}_{C}{}^{\dot{B}}+
\sigma^{b\dag}{}_{\dot{A}}{}^{C}\sigma^{a}{}_{C}{}^{\dot{B}}=2\delta^{ab}\delta_{\dot{A}}{}^{\dot{B}}.
\end{eqnarray}
We use anti-symmetric matrices
\begin{eqnarray}
\epsilon_{AB}=-\epsilon^{AB}=\begin{pmatrix} 0&-1 \\ 1&0
\end{pmatrix}\;
\;\;{\rm and}\;\;\;
\epsilon_{\dot{A}\dot{B}}=-\epsilon^{\dot{A}\dot{B}}=\begin{pmatrix}
0&1 \\-1& 0
\end{pmatrix}
\end{eqnarray}
to raise or lower un-dotted and dotted indices, respectively. For
example,
$\sigma^{a\dag\dot{A}B}=\epsilon^{\dot{A}\dot{B}}\sigma^{a\dag}{}_{\dot{B}}{}^{B}$
and $\sigma^{aB\dot{A}}=\epsilon^{BC}\sigma^{a}{}_{C}{}^{\dot{A}}$.
The sigma matrix $\sigma^a$ satisfies a reality condition
\begin{equation}\label{RC4}
\sigma^{a\dag}{}_{\dot{A}}{}^{B}=-\epsilon^{BC}\epsilon_{\dot{A}\dot{B}}\sigma^a{}_{C}{}^{\dot{B}},\quad
{\rm or} \quad\sigma^{a\dag\dot{A}B}=-\sigma^{aB\dot{A}}.
\end{equation}
The antisymmetric matrix $\epsilon_{AB}$ satisfies an important
identity
\begin{equation}
\epsilon_{AB}\epsilon^{CD}=-(\delta_A{}^C\delta_B{}^{D}-\delta_A{}^D\delta_B{}^{C}),
\end{equation}
and $\epsilon_{\dot{A}\dot{B}}$ satisfies a similar identity.

The parameter for the $\CN=4$ supersymmetry transformations is
defined as $\epsilon^{A\dot{B}}=\ep_a\s^{aA\dot{B}}$.

\section{The Commutation Relations of Superalgebras}\label{superalgebras}
These commutation relations of superalgebras are adopted from our previous paper \cite{ChenWu4}.
\subsection{$U(M|N)$}\label{cmrunum}
The commutation relations of $U(M|N)$ are given by
\e\label{unum}\nonumber
&&[M_{\bu}{}^{\bv},M_{\bw}{}^{\bt}]=\d_{\bw}{}^{\bv}M_{\bu}{}^{\bt}-\d_{\bu}{}^{\bt}M_{\bw}{}^{\bv},\quad [M_\pri{}^{\pj},M_\pk{}^\pl]=\d_\pk{}^\pj M_\pri{}^\pl-\d_\pri{}^\pl M_{\pk}{}^\pj\nonumber\\
&&[M_{\bu}{}^{\bv},Q_{\bw}{}^\pk]=\d_{\bw}{}^{\bv}Q_{\bu}{}^\pk,\quad [M_{\bu}{}^{\bv},\bar Q_\pk{}^{\bw}]=-\d_{\bu}{}^{\bw}\bar Q_\pk{}^{\bv},\nonumber\\
&&[M_\pri{}^{\pj},Q_{\bw}{}^\pk]=-\d_\pri{}^\pk Q_{\bw}{}^\pj,\quad [M_\pri{}^{\pj},\bar Q_\pk{}^{\bw}]=\d_\pk{}^\pj \bar Q_\pri{}^{\bw}\nonumber\\
&& \{Q_{\bu}{}^\pri,\bar Q_\pj{}^{\bv}\}=k(\d_\pj{}^\pri M_{\bu}{}^{\bv}+\d_{\bu}{}^{\bv}M_{\pj}{}^\pri),
\ee
where $Q_{\bu}{}^\pri$ carries a $U(M)$ fundamental index $\bu=1,\cdots,M$ and a $U(N)$ anti-fundamental index $\pri=1,\cdots,N$.
Here we have
\e\label{dcmpq}
Q_\pa=\begin{pmatrix} \bar Q_\pri{}^{\bu}\\-Q_{\bu}{}^\pri\end{pmatrix}=\bar Q_\pri{}^{\bu}\d_{1\a}-Q_{\bu}{}^\pri\d_{2\a},
\ee
 In the second equation of (\ref{dcmpq}), we have introduced a ``spin up" spinor $\chi_{1\alpha}$ and a ``spin
down" spinor $\chi_{2\alpha}$, i.e., \footnote{Here the index $\a$
is \emph{not} a spacetime spinor index. We hope this will not
cause any confusion.}
\begin{eqnarray}
\chi_{1\alpha}=\begin{pmatrix} 1 \\ 0
\end{pmatrix}=\d_{1\a}\;
\;\;{\rm and}\;\;\; \chi_{2\alpha}=\begin{pmatrix} 0 \\ 1
\end{pmatrix}=\d_{2\a}.
\end{eqnarray}
And the anti-symmetric tensor $\om_{ab}$ and its inverse read
\begin{equation}\label{dcmo}
\omega_{\pa\pb}=\begin{pmatrix} 0 & \d_{\bv}{}^{\bu}\d_{\pri}{}^\pj \\
-\d^{\bv}{}_{\bu}\d^{\pri}{}_\pj & 0
\end{pmatrix},\quad \omega^{\pb\pc}=\begin{pmatrix} 0 & -\d^{\pj}{}_\pk\d^{\bw}{}_{\bv} \\
\d_{\pj}{}^\pk\d_{\bw}{}^{\bv} & 0
\end{pmatrix}.
\end{equation}
With (\ref{dcmpq}) and (\ref{dcmo}), the superalgebra (\ref{unum})
takes the form of (\ref{slie2}) or (\ref{slie3}).

\subsection{$OSp(M|2N)$}\label{cmrsp22n}
The super Lie algebra $OSp(M|2N)$ reads
\begin{eqnarray}\label{OSp}
&&[M_{\bar i\bar j},M_{\bar k\bar
l}]=\d_{\bj\bk}M_{\bi\bl}-\d_{\bi\bk}M_{\bj\bl}+\d_{\bi\bl}M_{\bj\bk}
-\d_{\bj\bl}M_{\bi\bk},\nonumber\\
&&[M_{\hi\hj},M_{\hk\hl}]=\omega_{\hj\hk}M_{\hi\hl}+\omega_{\hi\hk}M_{\hj\hl}
+\omega_{\hi\hl}M_{\hj\hk}+\omega_{\hj\hl}M_{\hi\hk},\nonumber\\
&&[M_{\bi\bj},Q_{\bk\hk}]=\d_{\bj\bk}Q_{\bi\hk}-\d_{\bi\bk}Q_{\bj\hk},\nonumber\\
&&[M_{\hi\hj},Q_{\bk\hk}]=\omega_{\hj\hk}Q_{\bk\hi}+\omega_{\hi\hk}Q_{\bk\hj},\nonumber\\
&&\{Q_{\bi\hi},Q_{\bj\hj}\}=k(\omega_{\hi\hj}M_{\bi\bj}+\d_{\bi\bj}M_{\hi\hj}),
\end{eqnarray}
where $\bi=1,\cdots,M$ is an $SO(M)$ fundamental index, and
$\hi=1,\cdots, 2N$ an $Sp(2N)$ fundamental index. Here we have
\begin{equation}
Q_a=Q_{\bi\hi} \quad {\rm and}\quad
\omega_{ab}=\omega_{\bi\hi,\bj\hj}=\d_{\bi\bj}\omega_{\hi\hj}.
\end{equation}
Now the superalgebra (\ref{cmrsp22n}) also takes the form of (\ref{slie2}) or (\ref{slie3}).

\end{document}